\documentstyle[twocolumn,psfig,rotating]{mn}

\def\lsim{~\rlap{$<$}{\lower 1.0ex\hbox{$\sim$}}}
\def\bsim{~\rlap{$>$}{\lower 1.0ex\hbox{$\sim$}}}
\def\ssp{\,{\rm s^{-1}}}
\def\kms{\ {\rm km\,s^{-1}}}
\def\skm{\ {\rm s\,km^{-1}}}
\def\kmsmpc{\ {\rm km\,s^{-1}\,Mpc^{-1}}}
\def\hmpc{\ {\rm {\it h}^{-1}Mpc}}
\def\mdh{\ {\rm M_\odot/{\it h}}}
\def\hmmpc{\ {\rm {\it h}Mpc^{-1}}}
\def\cmm{\ {\rm cm^{-2}}}
\def\ccm{\ {\rm cm^{2}}}
\def\kel{\,{\rm K}}
\def\dd{{\rm d}}
\def\ln{{\rm ln}}
\def\rarrow{\rightarrow}

\def\la{\langle}
\def\ra{\rangle}
\def\vp{{\bf p}}
\def\vx{{\bf x}}
\def\vy{{\bf y}}
\def\vk{{\bf k}}
\def\vr{{\bf r}}
\def\dgnl{\delta_{\rm g}}
\def\dmnl{\delta_{\rm m}}
\def\vgnl{{\bf v}_{\rm g}}
\def\vmnl{{\bf v}_{\rm m}}
\def\sgl{\sigma_{\rm L}}
\def\sgnl{\sigma_{\rm NL}}
\def\xj{x_{\rm J}}

\def\xf{x_{\rm F}}
\def\xhf{\hat{x}_{\rm F}}
\def\kj{k_{\rm J}}

\def\kf{k_{\rm F}}
\def\khf{\hat{k}_{\rm F}}

\def\dhi{n_{_{\rm HI}}}

\def\tg{T_{\rm g}}
\def\thg{\hat{T}_{\rm g}}
\def\thgg{\hat{T}_4\!}
\def\dhhi{{\hat n}_{_{\rm HI}}}
\def\nhi{N_{_{\rm HI}}}
\def\dmk{\Delta_{\rm m}^2(k)}
\def\dfk{\Delta_{\rm F}^2(k)}
\def\dlk{\Delta_{\rm L}^2(k)}

\def\om#1{\Omega_{#1}}
\def\pmb#1{\setbox0=\hbox{#1}%
\kern-.025em\copy0\kern-\wd0
\kern.05em\copy0\kern-\wd0
\kern-.025em\raise.0433em\box0}

\def\etal{{\it et al.\ }}
\newcommand{\op}{Ly$\alpha$\ }
\newcommand{\hi}{\mbox{H{\scriptsize I}}}

\begin{document}
\title[] {Joint modeling of the probability distribution and power 
spectrum of the \op forest~: comparison with observations at $z=3$} 
\author[Desjacques {\it et al.}]{Vincent Desjacques$^1$ and Adi 
Nusser$^{1,2}$ \\ 
$^1$The Physics Department and the Asher Space Research Institute, 
Technion, Haifa 32000, Israel \\
$^2$ The Institute for Advanced Study, School of Natural Sciences, 
Einstein Drive, Princeton, NJ 08540  \\
Email~: dvince@physics.technion.ac.il, adi@physics.technion.ac.il\\} 
\maketitle

\begin{abstract}
We present results of joint modeling of the probability distribution
function (PDF) and the one-dimensional power spectrum (PS) of the \op
forest flux decrement.  The sensitivity of these statistical measures
to the shape and amplitude  of the linear matter power spectrum is
investigated using N-body simulations  of two variants of the
$\Lambda$CDM cosmology. In the first model, the linear power spectrum
has a scale-invariant spectral  index $n_s=1$, whereas in the second,
it has a negative running index  (RSI), $\dd n/\dd\ln k<0$. We
generate mock catalogs of QSO spectra, and  compare their statistical
properties to those of the observations at  $z=3$. We perform a joint
fit of the  power spectrum and  the PDF. A scale-invariant model with
$\sigma_8=0.9$ matches well  the data if the mean IGM temperature is
$T\lsim 1.5\times 10^4\kel$.  For higher temperature, it tends to
overestimate the flux power spectrum  over scales $k\lsim 0.01\skm$.
The discrepancy is less severe when the PS alone is  fitted. However,
models matching the PS alone do not yield a  good fit to the PDF.  A
joint analysis of the flux PS and PDF tightens the constraints on the
model parameters and reduces systematic biases.  The RSI model is
consistent with  the observed PDF and PS only if the  temperature is
$T\bsim 2\times 10^4\kel$.  The best fit models reproduce the slope
and normalisation of the column  density distribution, irrespective of
the shape and amplitude of the linear power spectrum. They are also
consistent with  the observed  line-width distribution given the large
uncertainties. Our joint analysis suggests that $\sigma_8$ is likely
to be in the range 0.7-0.9 for a temperature 
$1\lsim T\lsim 2\times 10^4\kel$ and a reasonable reionization history.
\end{abstract}


\section {Introduction}
\label{introduction}

Absorption features in the \op forest provide direct information on
the large scale distribution of neutral hydrogen in the highly ionized
intergalactic medium (Bahcall \& Salpeter 1965; Gunn \& Peterson 1965;
see e.g. Rauch 1998 for a review).  The \op forest is believed to
originate in a warm photoionized, fluctuating intergalactic medium
(IGM) which smoothly  traces the distribution of the dark matter.
This picture  is sustained by hydrodynamical simulations and
semi-analytical models which have succeeded in reproducing the
observed statistics of the absorption lines and transmitted flux
(e.g. McGill 1990; Bi 1993; Cen \etal 1994; Zhang, Anninos \&  Norman
1995; Petitjean, M\"ucket \& Kates 1995; Hernquist \etal 1996; Katz
\etal 1996; Miralda-Escud\'e \etal 1996; Hui, Gnedin \& Zhang 1997;
Theuns  \etal 1998; Schaye 2001).  The \op forest can therefore serve
as a probe the physical state of the IGM, and the underlying matter
distribution over a wide range  of scales and redshifts (e.g. Croft
\etal 1998; Nusser \& Haehnelt 2000;  Schaye \etal 2000).  Several
method have been proposed for  recovering  the shape  and amplitude of
the primordial power spectrum, $\dlk$ (e.g. Croft \etal 1998, 2002b;
McDonald \etal 2000; McDonald 2003). The current methods rely either
on an inversion of the one-dimensional flux power spectrum  $\dfk$
(e.g. Croft \etal 1999, 2002b; McDonald 2003), or on a forward
comparison between the data and the simulations (e.g. Zaldarriaga, Hui
\& Tegmark 2001; McDonald \etal 2004b). The nonlinear relation between
the matter and flux power spectra is calibrated by means of detailed
numerical simulations (e.g. Croft \etal 1998; McDonald 2003).  These
methods are fundamentally limited by continuum fitting errors, metal
contamination  and nonlinear corrections (e.g. Hui \etal 2001;
Zaldarriaga, Scoccimarro  \& Hui 2003; see also Gnedin \& Hamilton
2002). Furthermore, one has to marginalize over the 'nuisance'
parameters in order to place constraints  on the spectral index $n_s$
and the power spectrum normalisation amplitude $\sigma_8$.
Underestimating the mean flux level $\la F\ra$, for example, can
strongly  bias the results as it was shown by Seljak, McDonald \&
Makarov (2003),  and Viel, Haehnelt \& Springel (2004).  The most
recent analyses of the \op forest favour a  simple scale-invariant
model with $\sigma_8$ in the range 0.85-0.95 (McDonald \etal 2004b;
Viel, Haehnelt \& Springel 2004;  Viel, Weller \& Haehnelt 2004;
Seljak \etal 2004). However, one should emphasize that these studies
do not include in their analysis other statistics than the power
spectrum of the \op forest. A priori, a model which fits the  observed
power spectrum solely  will not necessarily agree with the observed
PDF, bispectrum etc. It is therefore prudent to examine how other
statistical measures of the forest constrain the cosmological model.

In this work, we follow a different strategy and assess the validity
of a cosmological model using three statistical measures of the \op
forest. These statistics are the one-dimensional  flux power spectrum,
$\dfk$, the flux probability distribution function (PDF), $P(F)$, and
the neutral hydrogen column density distribution
$f(\nhi)$. Furthermore,  we adopt the ``forward approach'' which
consists in comparing the  observed statistical measures with those
directly derived from N-body  simulations of the cosmological model
under consideration (see e.g.  Zaldarriaga, Hui \& Tegmark 2001). This
approach  has the advantage that  it does not require any inversion
procedure to obtain the linear mass  power spectrum.

Combining several statistical measures could significantly affect the
constraints on the shape and amplitude of the primordial power
spectrum as inferred from $\dfk$ alone. In this study, we will attempt
to match simultaneously the three  statistics mentioned above.  In
principle, higher order correlations such  as the bispectrum should
also be considered since they provide independent  information on the
cosmology (e.g. Mandelbaum \etal 2003). However,   the  \op bispectrum
as computed from high-resolution QSO spectra is still too noisy to
provide significant constraints  on the cosmological model (e.g. Viel
\etal 2004a).  To explore a parameter space as large as possible, we
will implement a  simple semi-analytical model which reproduces many
properties of the \op  absorbers (e.g. Bi, B\"orner \& Chu 1992;
Reisenegger \& Miralda-Escud\'e  1995; Bi \& Davidsen 1997; Gnedin \&
Hui 1998). We will generate a wide sample of mock catalogs and
constrain the model parameters by comparing their statistical
properties to the observations of McDonald \etal (2000).

The paper is organized as follows. In Section \S\ref{data} we briefly
review the observational data we aim at fitting in this analysis.  In
\S\ref{igm},  we detail the method used to compute synthetic spectra
from pure dark matter (DM) simulations.  The comparison between
simulation and observations is done in \S\ref{compare}. We discuss
our results in \S\ref{discussion}.

\section{Observational data}
\label{data}

In this paper, we will compare the simulated statistics to the flux
power spectrum  and PDF measured by McDonald \etal (2000) (hereafter
M00), which were  obtained from a sample of eight high-resolution QSO
spectra.  Regarding the flux power spectrum, we consider the data
points in the  range $0.005<k<0.04\skm$. The lower limit $k=0.005\skm$
is set by the size of our simulations
(cf. Section~\S\ref{nbody}). Note that it is larger than the
characteristic   wavenumber $k\sim 0.003\skm$ below which continuum
fitting errors are expected to become important (e.g. Hui \etal
2001). The upper limit  $k=0.05\skm$ is chosen because of concern
about metal contamination on smaller scales (e.g. M00; Kim \etal
2004).  The observed flux PDF is very sensitive to continuum fitting,
especially in the high transmissivity tail. We therefore exclude the
data points with $F\geq 0.8$ from the analysis to avoid dealing with
continuum fitting  error. In addition, we discard also the measurement
at $F=0$ since it is  strongly affected by noise (e.g. Rauch \etal
1997).  Finally, we emphasize that we work with the power spectrum of
the transmitted flux $F$, $\dfk$, which  differs from that of the flux
density contrast,  $\delta_{\rm F}\equiv\left(F/\la F\ra -1\right)$,
by a multiplicative  factor $\la F\ra^2$. $\dfk$ has the advantage
that it does not depend explicitly on the mean flux level $\la F\ra$.
	
\begin{table}
\caption{The main parameters of the $\Lambda$CDM simulations 
considered in the present paper. The normalisation of the RSI 
simulations follows from the results of Spergel \etal (2003).
The spectral index $n_s$ and its derivative $\dd\ln n_s/\dd\ln k$
are given at $k=0.07\hmmpc$. The mass of a dark matter particle is 
about $8\times 10^7\mdh$ in all the simulations.}
\vspace{1mm}
\begin{center}
\begin{tabular}{ccccccc} \hline
& $\om{m}$ & $\om{\Lambda}$ & h & $\sigma_8$ & $n_s$ &
$\dd\ln n_s/\dd\ln k $ \\ \hline
S1 and S2  & 0.30 & 0.70 & 0.70 & 0.90 & 1 & 0 \\
R1 and R2  & 0.31 & 0.69 & 0.71 & 0.84 & 0.93 & -0.03 \\
\hline\hline
\end{tabular}
\end{center}
\label{table1}
\end{table}

\begin{figure*}
\resizebox{0.45\textwidth}{!}{\includegraphics{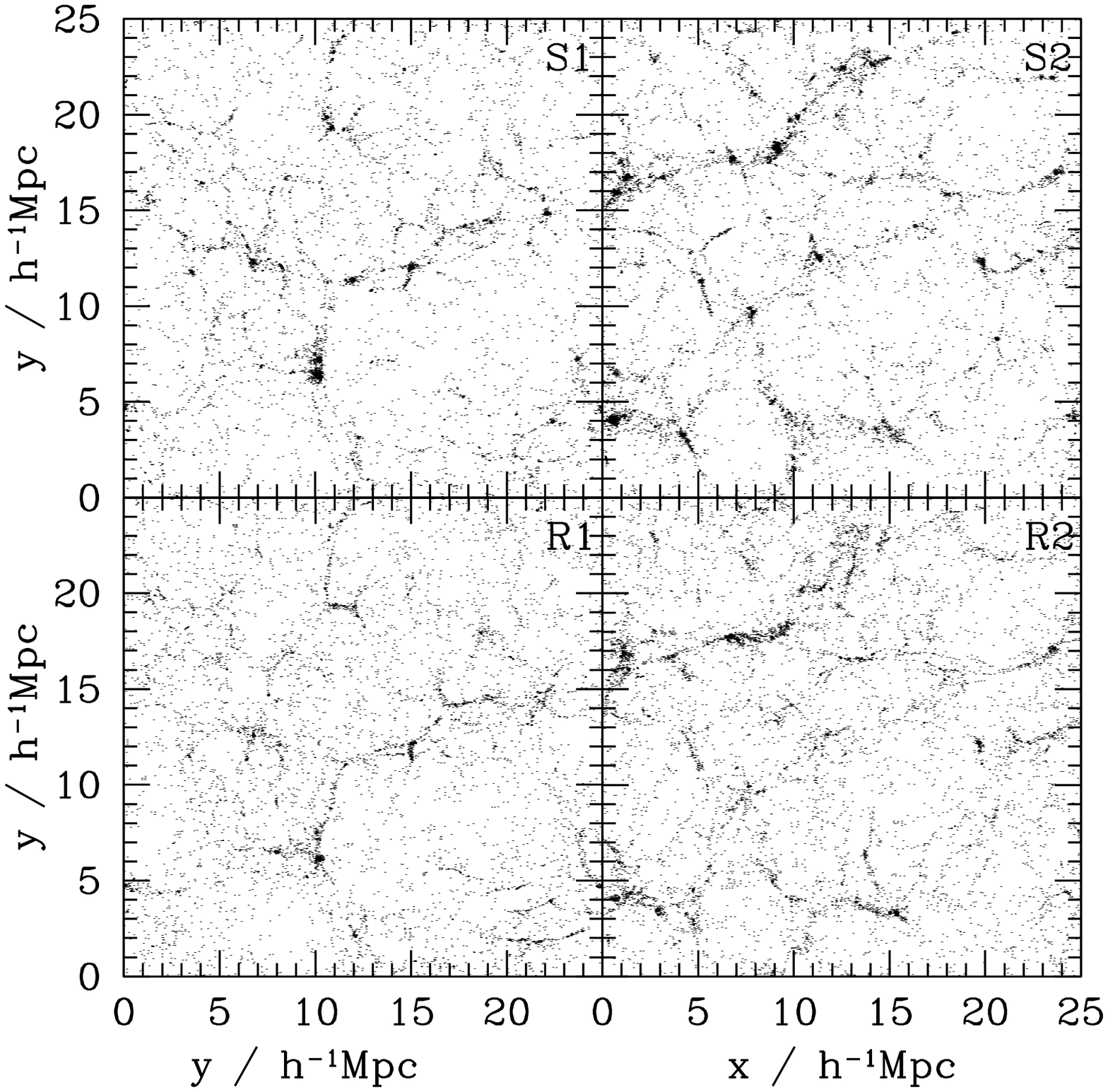}}
\resizebox{0.45\textwidth}{!}{\includegraphics{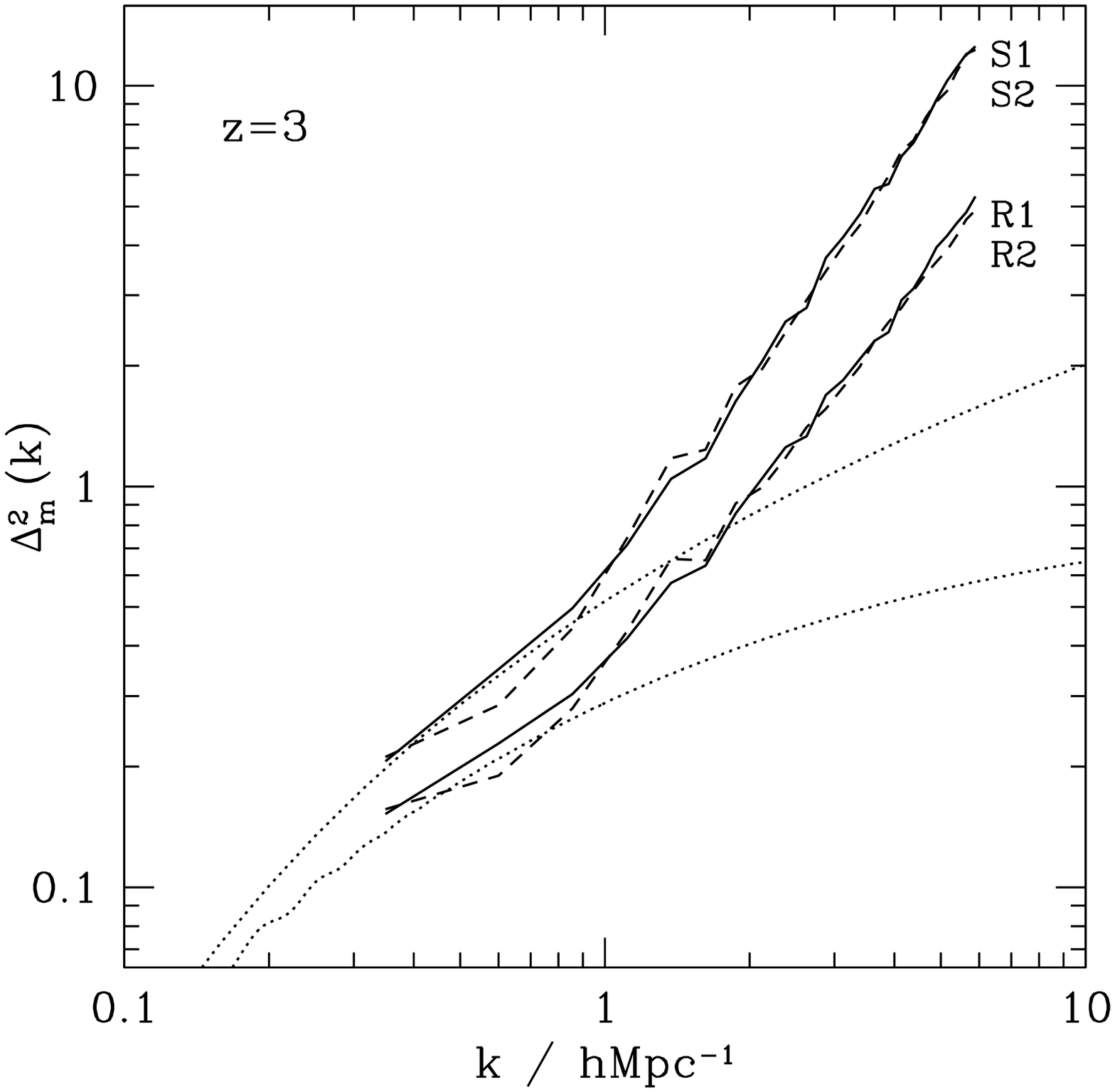}}
\caption{
{\it Left panel}~: The particle distribution in a slice of thickness 
$\Delta h=0.05\hmpc$ through the various $\Lambda$CDM simulations at  
$z=3$~: the scale-invariant realisations S1 and S2 (top panels), and 
the RSI realisations R1 and R2 (bottom panels).
{\it Right panel}~: The corresponding 3D power spectra $\dmk$ computed 
at $z=3$ from the DM particle distribution. The dashed curves show
$\dmk$ for the S2 and R2 simulations. The dotted curves show 
the linear power spectrum used to seed the fluctuations.}
\label{fig1}
\end{figure*}

\section{Modelling the \op forest}
\label{igm}

In this Section, we will first describe the DM simulations in detail. 
We will then briefly review the basic ingredients relevant to create
realistic \op spectra from simulations of DM only. 

\subsection{The N-body simulations}
\label{nbody}

We run DM simulations of a $\Lambda$CDM cosmology with the N-body code
GADGET (Springel, Yoshida \& White 2001). Pure DM simulations are more
suitable to our purposes since they allow us to explore  a larger
parameter space than full hydrodynamical simulations.  To reproduce
the \op forest as seen in QSO spectra and, at the same time, resolve
typical absorption systems, a large volume together with a high
resolution are requested (e.g. Theuns \etal 1998; Bryan \etal 1999).
For this reason, each simulation evolves 256$^3$ particles of mass
$m\sim 10^8\mdh$ in a periodic box of size 25$\hmpc$. Since cosmic
variance error on, e.g., the optical depth normalisation (at constant
mean flux) can be relatively large when estimated from such a small
volume (e.g. Croft \etal 2002b), we have run two random
realisations  of each of the model we consider in the present
paper. In the  simulations which we denote S1 and S2, the initial
power spectrum is  scale-invariant, $n_s=1$, and was obtained from 
fitting formulae of Eisenstein \& Hu (1999).
In the simulations R1 and R2, the initial power spectrum is that of
the running spectral index model (RSI, see e.g. Spergel 2003).  The
present-day clustering  amplitude in these simulations,
$\sigma_8=0.84$, is slightly lower  than that of S1 and S2, for which
$\sigma_8=0.9$. The initial particle positions were computed using a
parallel code kindly provided by Volker Springel.  The cosmological
parameters are summarized in Table~\ref{table1}.  In the left panel of
Fig.~\ref{fig1}, we show a slice of thickness  $\Delta h=0.05\hmpc$
extracted from the four simulations at redshift  $z=3$. Note that the
simulations S1 and R1 (S2 and R2) have the same  random seed. 

To quantify to which extent the various simulations differ, we plot in
Fig.~\ref{fig1} the dimensionless 3D matter power spectrum $\dmk$ at
$z=3$ for all the simulations. It was obtained  by sampling the DM
particles onto a 512$^3$ grid using a cloud-in-cell method, and taking
the  Fourier transform by means of a FFT routine. As we can see, R1
and R2 exhibit much less power than S1 and S2. On scale $k\sim
3\hmmpc$, which corresponds to the pivot point in the flux power
spectrum (cf.  Figure~\ref{fig4}), $\dmk$  in the scale-invariant
simulations is about twice as large as is in the RSI simulations.
Cosmic variance becomes important on scale $k\lsim 1\hmmpc$, where the
matter power spectrum computed from the S1 (R1) realisation is larger
than that of S2 (R2) by about 10-20\%.  We also show the linear power
spectra which we use to seed fluctuations in  the initial DM
distribution as dotted curves. The largest  scale of the simulations,
$k\lsim 1\hmmpc$, are still in the linear  regime at $z=3$. Note also
that,  on scale $k\lsim 0.1\hmmpc$, the power spectrum of the RSI
model is very close to that of the $n_s=1$ model.

\begin{figure*}
\resizebox{1\textwidth}{!}{\includegraphics{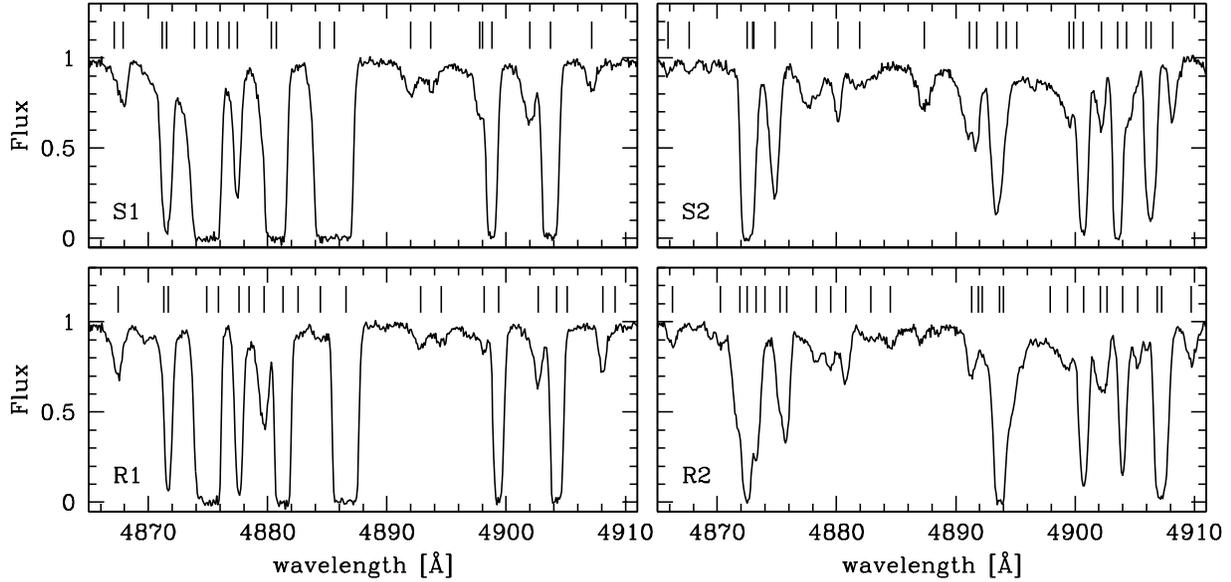}}
\caption{Synthetic spectra extracted from each of the simulations at 
$z=3$. The vertical bars above the spectra show the position of the lines 
with fitted column densities $\nhi$ exceeding $10^{12.5}\ccm$. The comoving 
length of each spectrum is $L=25\hmpc$, and corresponds to a redshift 
interval $\Delta z\sim 0.04$ at $z=3$.}
\label{fig2}
\end{figure*}

\subsection{Generating mock catalogs of QSO spectra}

\subsubsection{The low density IGM}
\label{igm}

Semi-analytical models of the \op forest exploit the tight
relationship  between the gas and the underlying DM distribution (e.g.
Bi 1993; Bi \& Davidsen 1997; Petitjean \etal 1995; Gnedin \& Hui
1998;  Croft \etal 1998).  In this approach, the gas temperature
$\tg$, and the neutral hydrogen density $\dhi$ are computed using
tight power-law relations obtained  from full hydrodynamical
simulations (e.g. Katz \etal 1996, Hui \&  Gnedin 1997, Theuns \etal
1998),
\begin{equation}
\tg=\thg\left(1+\dgnl\right)^{\gamma-1}~~~ \mbox{and}~~~
\dhi=\dhhi\left(1+\dgnl\right)^\alpha\;,
\label{lowigm}
\end{equation}
where the adiabatic index $\gamma$ is in the range $1-1.62$, and
$\alpha=2-0.7\left(\gamma-1\right)$. $\thg$ and $\dhhi$ are the
temperature and neutral hydrogen density at mean gas density
respectively.  These scaling relations hold for moderate gas
overdensities, i.e.  $\dgnl\lsim 50$~\footnote{When pressure smoothing
is properly accounted for, this approximation indeed breaks down at
much lower overdensities (Schaye 2001)}. For higher overdensities
however, we have to account  for line cooling. To this extent, we
follow Desjacques \etal (2004) and  take the gas temperature to be
$\tg=10^4\kel$ for $\delta\rho/\rho>50$ This cutoff ensures that the
Doppler parameter $b(\vx)$ is always  $b(\vx)=13\kms$ in the high
density regions of the simulation.

When radiative cooling is inefficient, it is reasonable to assume that
the gas density and velocity fields, $\dgnl$ and $\vgnl$, are smoothed
versions of their DM counterparts, $\dmnl$ and $\vmnl$. We obtain the
gas  density distribution by sampling the DM particles onto a 512$^3$
grid,  and smoothing the resulting density field with a Gaussian
filter  $W=\exp(-k^2/2\kf^2)$ of  characteristic length $1/\kf$. This
choice is motivated by the fact  that a Gaussian filter gives a good
fit to the gas fluctuations over a wide range of wavenumber 
(e.g. Gnedin \etal 2003).
Furthermore, we expect $\kf$ to depend on the physical state of the
IGM. However, since the relation between $\kf$ and $\thg$ depends
noticeably on the reionization history of the Universe (Gnedin \& Hui
1998), it is more  convenient to treat $\kf$ as a free parameter (see
e.g. Zaldarriaga, Hui \& Tegmark 2001). We discuss gas smoothing in
more detail in  Appendix~\S\ref{filtering}.

\subsubsection{The synthetic spectra}
\label{spectra}

Once we have smoothed the DM density and velocity fields on the
comoving scale $1/\kf$, the flux distribution depends only on the mean
flux $\la F\ra$,  the adiabatic index $\gamma$ and the mean IGM
temperature $\thgg$.  For any value taken by the four-dimensional
parameter vector $\vp$=($\kf$,$\la F\ra$,$\gamma$,$\thgg$), we
generate mock catalogs as follows.  For each of the simulations, we
randomly select 10$^4$ lines of sight (LOS)  of comoving length
$L=25\hmpc$. The optical  depth $\tau$ and the transmitted flux
$F=\exp(-\tau)$ are then computed  along each LOS in 512 pixels of
width $\Delta v\simeq 6\kms$ according to  the Gunn-Peterson
approximation (Gunn \& Peterson 1965). The resolution of the synthetic
spectra is somewhat larger than that of the M00 data, where it is
$\Delta v=2.5\kms$. In order to create realistic Keck spectra, we
should have computed the flux on pixels smaller than $2.5\kms$, smooth
the spectra with a Gaussian of resolution 6.6$\kms$ (FWHM), and then
resample them on pixels of $2.5\kms$. Unfortunately, FFT of
three-dimensional regular grids is still prohibitively slow for meshes
larger than $512^3$. Notwithstanding, we believe that the flux  power
spectrum and PDF will not be much affected. However, we should
emphasize that it might  not be the case of the line statistics, in
particular  the line width distribution.  In the last step, the value
of $\dhhi$ is adjusted such that the mean flux $\la F\ra$  of the
whole sample matches the desired value (e.g. Rauch \etal 1997). To
account for the noise in the observations, we add a uniform Gaussian
deviate  of dispersion $\sigma=0.02$ per interval of width
$\Delta\pi=2.5\kms$.

\begin{figure*}
\resizebox{0.68\textwidth}{!}{\includegraphics{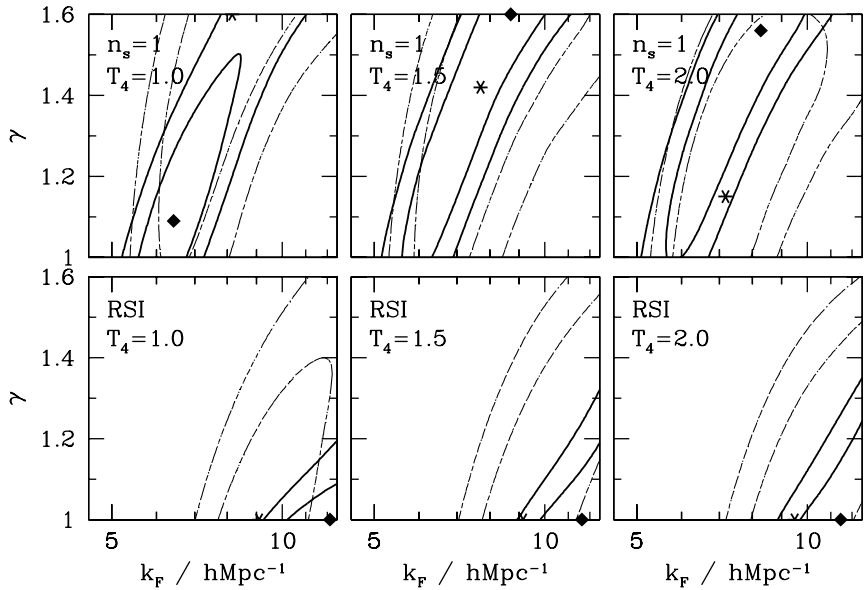}}
\caption{Constraints in the plane $\kf$ - $\gamma$ for a scale-invariant 
cosmology with $\sigma_8=0.9$ (top panels), and a RSI cosmology (bottom 
panels). The contours show the 1 and 2$\sigma$ confidence levels (assuming
a Gaussian likelihood). The mean IGM temperature is $\thgg=1$ (left panels), 
1.5 (middle panels) and 2 (right panels). 
The constraint arising from measurements of the flux PS solely is shown as 
dashed contours, while that arising from the combination of the flux PS and 
PDF is shown as solid contours. The stars and diamonds mark the best fit 
models obtained from the PS and PS+PDF data respectively.}
\label{fig3}
\end{figure*}

In Fig.~\ref{fig2}, we plot four synthetic spectra selected from each
of the simulations. The spectra were extracted from mock samples
which, for a given simulation, give an acceptable fit to the observed
statistics of  the \op forest (cf. Section~\S\ref{compare}). In
particular, the mean IGM temperature is respectively $\thgg=1.5$ and 2
for the scale-invariant (top panels) and RSI models (bottom panels).
Ticks above the spectra mark the \op absorption lines which were
identified with the spectral fitting program VPFIT (Carswell \etal
2003).  The lines are distinguished by their column density $\nhi$ (in
$\cmm$) and their width or Doppler parameter $b$ (in $\kms$). This
fitting technique  assumes that the fundamental shape of the lines is
a Voigt profile, which is a good approximation for column densities
$\nhi\lsim 10^{17}\cmm$ relevant to the \op forest.  Absorption
systems with column density $\nhi\leq 10^{12.5}\ccm$ are quite
uncertain and are not shown on the figure (e.g. Bi \& Davidsen
1997). The spectra extracted from R1 and R2 feature somewhat more
lines  than those extracted from S1 and S2. The reason is the larger
clustering  of the scale-invariant models. As a result, blending of
absorbers to form one strong line occurs more often, and cause the
number of lines identified  with VPFIT to be lower than that in the R1
and R2 models. Note that, in  practice, associated metal lines can
help fixing the number of  subcomponents. We will discuss the
properties of the simulated column density distribution in
Section~\S\ref{line}.

\section{Comparison with observations}
\label{compare}

In this Section, we compare the statistical properties of our mock
catalogs  to that of the observations. We perform a $\chi^2$
statistics for the   observed flux power spectrum and PDF. We also
compare the simulated  line column density and line-width
distributions to observational data for several models.

\subsection{The parameter  grid}
\label{grid}

For each value of our parameter vector 
$\vp$=($\kf$,$\la F\ra$,$\gamma$,$\thgg$), 
we generate mock catalogs of ten thousand lines of sight. We let the 
parameters assume the following values, 
\begin{eqnarray*}
\kf &=& 4.55,5,5.56,6.25,7.14,8.33,10,12.5 \\ \gamma &=&
1,1.1,1.2,1.3,1.4,1.5,1.6 \\ \thgg &=& 0.5,1,1.5,2,2.5 \\ \la F\ra &=&
0.60,0.61,0.62,\cdots,0.71,0.72,0.73
\end{eqnarray*}
The values of $\kf$ (in unit of $\hmmpc$) were chosen such that
$1/\kf$  uniformly spans the range $0.08-0.22\hmpc$.  Our discrete
grid thus contains 8$\times$7$\times$5$\times$14=3920 models.  For
each mock catalog, we then calculate the flux power spectrum and the
flux PDF, and average over the S1 and S2 (resp. R1 and R2)
realisations.  We then determine the goodness  of fit of any model in
the grid by computing a $\chi^2$ statistics from the difference
between the simulated  PS, PDF and the observational data.  We account
for measurements of the mean flux $\la F\ra$ by adding a term $(\la
F\ra-\bar{F})^2/\sigma_{\bar{F}}^2$ to the chi-squared, where
$\bar{F}$ and $\sigma_{\bar{F}}$ are the observed mean flux and error
bar at $z=3$.  Note that the number of flux PS  and PDF measurements
used in the analysis is 10+15=25.    Regarding the mean flux level,
we will hereafter adopt the default value of M00, $\bar{F}=0.684$ 
and $\sigma_{\bar{F}}=0.023$ (see Table~1 of their paper).  We
should however keep in mind that the presence of metal lines or strong
absorption systems can substantially affect the determination of $\la
F\ra$ (e.g. Schaye \etal 2003; Viel \etal 2004b).  We will discuss the
effect of changing the mean flux level in Section~\S\ref{errors}.  

The problem of finding the best fit models is equivalent of finding
the maximum of some hypersurface. Since the true maximum does not
generally lies exactly at a grid point, it is desirable to interpolate
over the different parameters. To this extent, we take advantage of
the smooth dependence of the flux PS and PDF on the parameter vector,
and use cubic spline interpolation. This method is known to work
substantially  better than a  simple multi-linear interpolation
(e.g. Tegmark \& Zaldarriaga 2000).  To marginalize over a subset of
the parameters, we fix their values to that of the best fit model of
the subspace, which we obtained from the spline interpolation.  To
check the accuracy of the interpolation, we computed the flux PS and
PDF from the simulations for several randomly selected models which do
not lie at a grid point. We found that the flux PS and PDF of the
spline-interpolated models differ from that of the 'true' models by
$\lsim 0.5$\% only in the range of scales considered in our analysis.

The uniformity of space along the line of sight implies that the
different $k-$modes of the flux PS are uncorrelated.  Uneven sampling
of the  forest along the line of sight or removal of some regions in
the data (such as saturated lines) will introduce covariance between
the modes. Yet the covariance matrix of the observed  flux power
spectrum is reasonably close to diagonal over the  scales relevant to
our analysis (Croft \etal 2002, Viel, Haehnelt \& Springel 2004).  We
will thus neglect correlations between data points in the flux PS. In
the case of the PDF, however, the data points are strongly correlated.
As a result, the $\chi^2$ distribution as computed from the diagonal
elements of the error matrix solely might differ noticeably from that
computed from the full error matrix (see e.g. M00; Gnedin \& Hamilton
2002).  The $\chi^2$ should therefore be computed from the full
covariance matrix (available at
http://www.astronomy.ohio-state.edu/\~jordi/lya).  This covariance
matrix reflects the random errors in the flux measurement and, more
importantly, the errors resulting  from the finite number of QSO
spectra. It was estimated  using a Monte-Carlo procedure based on a
variation of the  bootstrap method (Press \etal 1992; see Appendix B
of M00 for the details).  Including the full covariance matrix could
have a strong effect on the results, especially when the errors are
highly correlated as  it is certainly the case in the flux PDF. It is
therefore important  to assess the sensitivity of our analysis to the
details of the  covariance matrix. To this purpose, we will initially
present results  obtained by setting to zero the off-diagonal elements
of the covariance matrix. In Section~\S\ref{matrix}, we will repeat
the analysis using the full covariance matrix of the flux PDF, and
discuss to which extent the inclusion of the covariance matrix affects
the inferred constraints.

Since the spectral fitting program is relatively slow, we have  not
included the line distribution in the $\chi^2$ statistics. We will
compute the line statistics only for models which best fit the  PS 
and PDF (cf. Section~\S\ref{line}).

\begin{figure*}
\resizebox{0.72\textwidth}{!}{\includegraphics{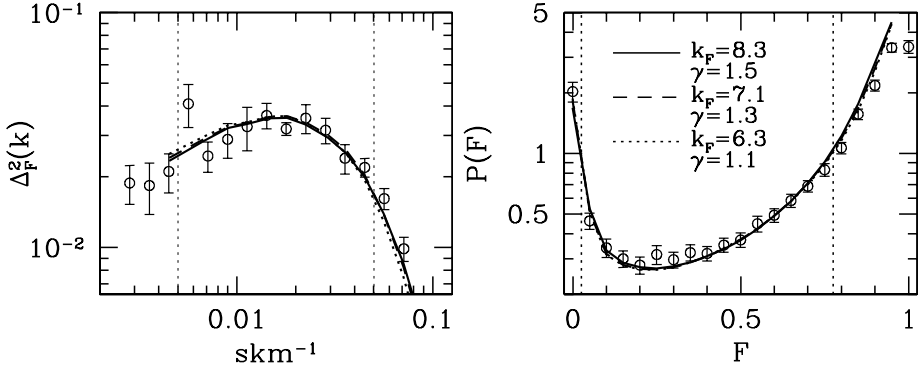}}
\caption{Three scale-invariant models of normalisation $\sigma_8=0.9$
which give an acceptable fit to the  observed flux PS and PDF. The
mean IGM temperature is $\thgg=1.5$. The reduced chi-squared are
$\chi^2$=23.5, 23.2 and 22.8 (for 23 degrees of freedom) for
$\gamma=1.5$, 1.3 and 1.1 respectively. Only the data points  which
lie between the two vertical lines were used to compute the $\chi^2$
statistics.}
\label{fig4}
\end{figure*}

\begin{figure*}
\resizebox{0.68\textwidth}{!}{\includegraphics{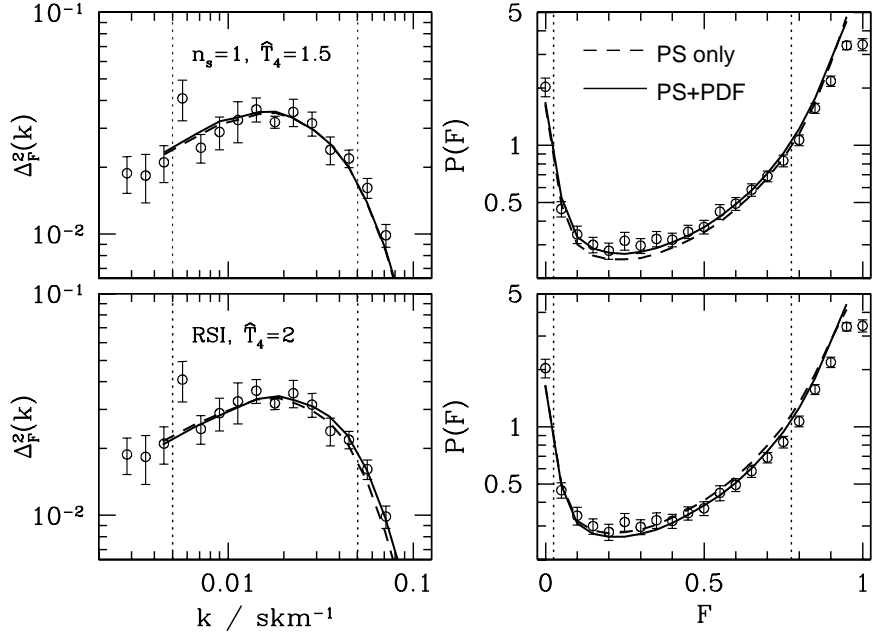}}
\caption{A comparison between models which best fit the PS data alone
(dashed curves), and models which best fit the PS+PDF data (solid
curves).  They are shown for the $\sigma_8=0.9$, scale-invariant 
cosmology (top panels), and for the RSI cosmology (bottom panels),
for a fixed temperature $\thgg=1.5$ and 2 respectively. For the PS 
data only, the best fit values of the other parameters 
are $(\kf,\la F\ra,\gamma)$=(7.7,0.71,1.42) and (9.5,0.69,1) in the 
$n_s=1$ and RSI models respectively, whereas with the PS+PDF data, 
they are (8.7,0.69,1.6) and (11.5,0.70,1).}
\label{fig5}
\end{figure*}

\subsection{The flux power spectrum and probability distribution}
\label{pspdf}

\subsubsection{Constraint in the $\kf$-$\gamma$ plane}
\label{plane}

\begin{figure*}
\resizebox{0.68\textwidth}{!}{\includegraphics{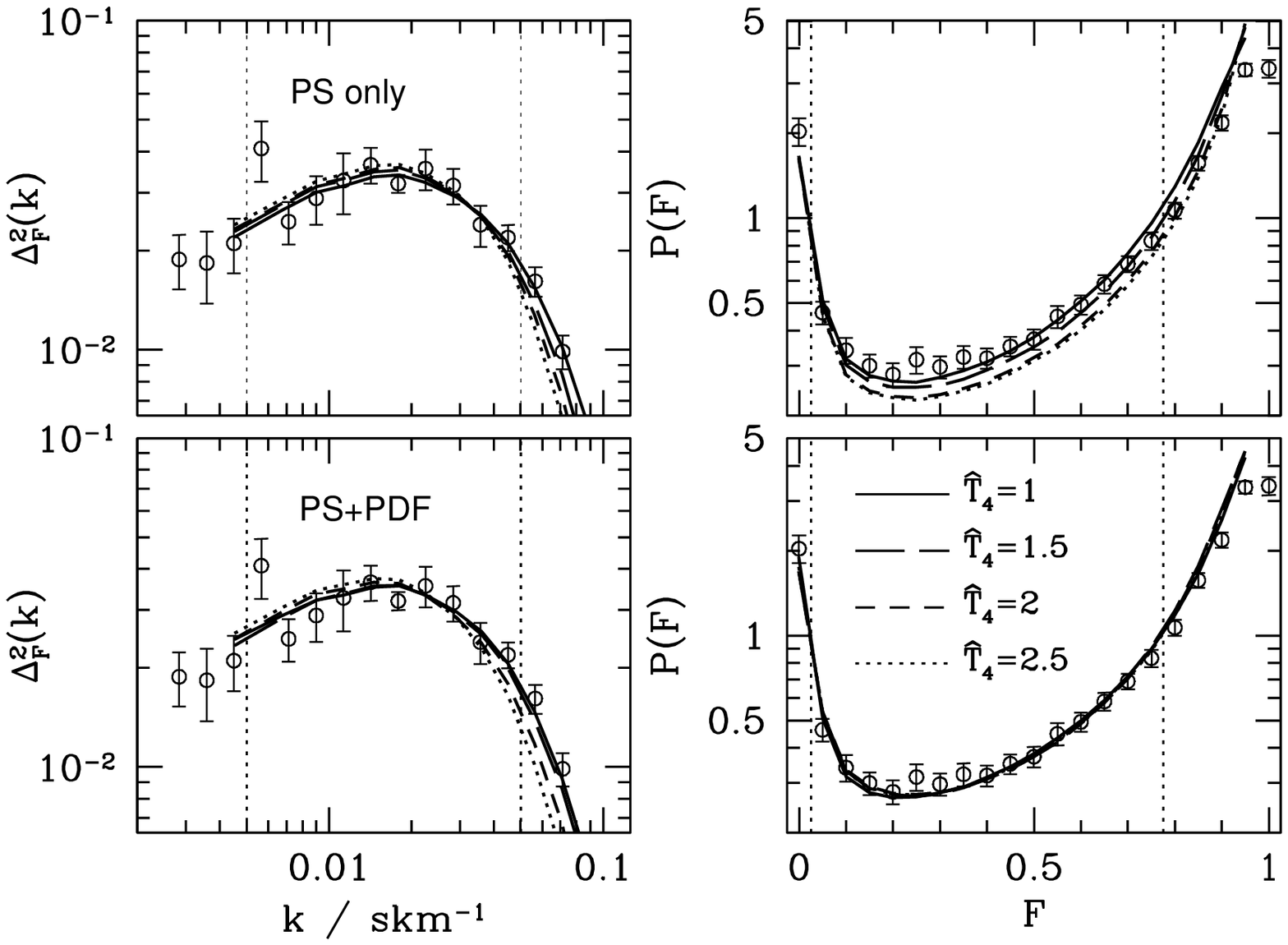}}
\caption{Scale-invariant models of rms fluctuation $\sigma_8=0.9$
which  fit best the PS data solely (top panels), and the PS+PDF data
(bottom panels) as a function of the mean IGM temperature. The
temperature has a fixed $\thgg=1$ (solid), 1.5 (long-dashed),  2 
(short-dashed) and 2.5 (dotted curves), while the other parameters
were varied. The best fit parameter values obtained with the PS+PDF 
data are listed in  Table~\ref{table2}.}
\label{fig6}
\end{figure*}

Fig.~\ref{fig3} shows  the constraints in the plane $\kf$ - $\gamma$ for
the scale-invariant cosmology with  $\sigma_8=0.9$ (top
panels) and for the RSI cosmology (bottom panels). The mean IGM
temperature is $\thgg=1$ (left panels), 1.5 (middle panels) and 2
(right  panels).  The contours show the 68.3\% and 95.5\% confidence
levels (assuming a  Gaussian distribution for the likelihood) obtained
from the flux power spectrum alone (dashed curves), and from a
combination of the flux PS and PDF data (solid curves). There is
obviously a degeneracy between the filtering wavenumber $\kf$ and  the
adiabatic index $\gamma$.  For illustration, we plot in
Fig.~\ref{fig4} the PS and PDF of three  scale-invariant models which
fit the PS and PDF data at the 1$\sigma$  level. The IGM temperature
is $\thgg=1.5$.  The models are plotted together with the measurements
of M00, shown as empty symbols. As we can see, although these  models
differ in the value of $\kf$ and $\gamma$, they all yield very similar
power spectra and PDFs.  To understand the origin of this degeneracy,
note that increasing  $\gamma$ generally reduces the width of the weak
lines ($\tau\lsim 1$) relative to that of the strong lines ($\tau\bsim
1$). This  follows from the fact that the low column density \op
forest arises in gas of low overdensity. 
Consequently, increasing $\gamma$ (at constant mean flux) amounts to
decreasing  $\dfk$ on small scale, and thereby mimics the effect of a
smaller $\kf$.  In the PDF, increasing $\gamma$ enhances the fraction
of transmissivity  pixels in the range $F\bsim 0.5$ presumably
because, at constant mean  flux, the lower optical depth normalisation
and the reduction in the  width of the weak lines conspire to increase
the fraction of high  transmissivity pixels.  As a result, the
combination of the PS and PDF cannot break the degeneracy since both
statistics exhibit a  similar dependence on $\kf$ and $\gamma$. In
reality however, one would  expect the filtering length to depend on
$\thg$ and $\gamma$  (see Appendix~\S\ref{filtering}) . Therefore,
additional assumptions on the reionization history of the Universe
could break this degeneracy.  In this respect, the observed line-width
distribution suggests that, around $z=3$, there is a sharp increase in
$\thg$ together with a decrease  in $\gamma$ (Schaye \etal 2000;
Ricotti, Gnedin \& Shull 2000; see however  McDonald \etal
2001). However, the data are too noisy to constrain $\thgg$  and
$\gamma$ significantly.

\subsubsection{The best fit models}

In Fig.~\ref{fig3}, the stars and squares symbols  mark the models
which  best fit the PS and PS+PDF data, respectively. These models are
compared  in Fig.~\ref{fig5} to the M00 data.  The temperature of the
scale-invariant models was chosen to be $\thgg=1.5$, while the RSI
models have $\thgg=2$. This choice is motivated by the fact that, in a
scale-invariant cosmology with  $\sigma_8=0.9$, models with
$\thgg\lsim 1.5$ fit the data best whereas, in a RSI cosmology, the
best fit are obtained for $\thgg \bsim 2$ (cf. Table~\ref{table2}).
The RSI models match the observed $\dfk$ somewhat better than  the
scale-invariant models. However, they tend to overestimate the PDF  in
the range $F\bsim 0.6$, which roughly traces regions of gas
overdensity $\dgnl\lsim 0$. This follows from the relative lack of
small-scale power, which translates into larger values of the PDF in
the high transmissivity tail.  Consequently, matching the flux PDF in
a RSI cosmology requires a low filtering length. Indeed, although the
RSI models which best fit the PS  data alone have $\kf\lsim 10\hmmpc$,
those which best fit the PS+PDF data  have $\kf\bsim 11\hmmpc$, close
to the largest value assumed by $\kf$ in  our parameter grid. Since
the Nyquist frequency of the grid is  $k_{\rm Ny}\approx 10.2\hmmpc$,
these models  are certainly affected by numerical resolution. However,
numerical  resolution also contributes to the smoothing
(e.g. Zaldarriaga, Hui \&  Tegmark 2001), and $\kf=11\hmmpc$ is merely
a lower limit.  Note that we have $\kf\sim 6-9\hmmpc$ for the best
scale-invariants  models, a filtering scale which is robust to
resolution issues.

Fig.~\ref{fig5} also clearly demonstrates that models which match best
the PS data  alone do not yield a good fit of the PDF. The constraints
inferred from measurements of $\dfk$ alone might thus be significantly
biased.  It is therefore important to combine the PS and PDF
statistics to  ensure that both are correctly reproduced in the
simulation.  To quantify the effect of adding the PDF, we plot in
Fig.~\ref{fig6}  the flux power spectrum and PDF of scale-invariant
models for several values of $\thg$. Models which fit best the PS and
the PS+PDF data are shown in the top and bottom panels,
respectively. When the PS data alone are fitted , the agreement
between the observed and simulated flux PDF worsens with increasing
temperature.  For $\thgg\bsim 2$, the disagreement is very severe. A
comparison  between the top and bottom left panels reveals that
including the PDF  in the $\chi^2$ statistics increases the amplitude
of flux power spectrum  by an amount of $\sim$10\% on scale $k\lsim
0.01\skm$. Increasing the  temperature produces a similar but weaker
effect~: the large-scale  amplitude of $\dfk$ decreases by  about 5\%
when the temperature increases from $\thgg=1$ to 2. Note that the
strength of this effect is consistent with findings from
hydrodynamical simulations (e.g. Viel, Haehnelt \& Springel 2004).
Combining the PS and PDF data could therefore reduce the best fit
value  of $\sigma_8$ by about 10\%. We also expect a degeneracy
between the $\sigma_{8}$ and the mean IGM temperature. The
data could be equally matched with a lower  $\sigma_8$ and a larger
IGM temperature. We will discuss this issue in more detail in
Section~\S\ref{sigma8}.

\begin{table}
\caption{Parameter values of scale-invariant and RSI models which best
fit the PS+PDF data, for a mean IGM temperature $\thgg=1$, 1.5, 2 and
2.5. The filtering $\kf$ is in unit of $\hmmpc$. The last column gives
the chi-squared for 23 degrees of freedom. Note that, since we spline 
interpolate over the parameters, the best fit values do not necessarily 
lie at a grid point.}
\vspace{1mm}
\begin{center}
\begin{tabular}{cc|ccc|c} \hline
model & $\thgg$ & $\la F\ra$ & $\kf$ & $\gamma$ & $\chi^2$ \\ \hline
$n_s=1$, $\sigma_8=0.72$ & 1 & 0.69 & 7.2 & 1 & 21.4 \\
                      & 1.5 & 0.70 & 11.1 & 1.6 & 18.9 \\
                      & 2 & 0.70 & 10.2 & 1.53 & 21.0 \\
                      & 2.5 & 0.70 & 8.8 & 1.37 & 24.9 \\
$n_s=1$, $\sigma_8=0.82$ & 1 & 0.69 & 6.6 & 1 & 20.9 \\
                      & 1.5 & 0.69 & 9.7 & 1.6 & 21.4 \\
                      & 2 & 0.70 & 9.1 & 1.55 & 25.1 \\
                      & 2.5 & 0.70 & 8.1 & 1.42 & 30.4 \\
$n_s=1$, $\sigma_8=0.9$ & 1 & 0.69 & 6.5 & 1.16 & 21.3 \\
                      & 1.5 & 0.69 & 8.8 & 1.6 & 23.7 \\
                      & 2 & 0.69 & 8.2 & 1.55 & 28.4 \\
                      & 2.5 & 0.69 & 7.7 & 1.45 & 34.9 \\
$n_s=1$, $\sigma_8=1$ & 1 & 0.68 & 6.9 & 1.29 & 21.3 \\
                      & 1.5 & 0.69 & 8.0 & 1.57 & 25.4 \\
                      & 2 & 0.69 & 7.9 & 1.56 & 32.0 \\
                      & 2.5 & 0.69 & 7.0 & 1.42 & 39.9 \\
RSI                   & 1 & 0.7 & 12.1 & 1 & 31.3 \\
                      & 1.5 & 0.7 & 11.6 & 1 & 25.7 \\
                      & 2 & 0.7 & 11.5 & 1 & 22.2 \\   
                      & 2.5 & 0.7 & 12.5 & 1.07 & 20.6 \\
\hline\hline
\end{tabular}
\end{center}
\label{table2}
\end{table}

The photoionisation rate $\Gamma_{-12}$ of the neutral hydrogen (in
unit of $10^{-12}\ssp$) can be estimated from the optical depth
normalisation $\tau_0$, which can be written as
\begin{equation}
\tau_0=2.31\,{\cal A}(z)\,\thgg^{-0.7}\left(\frac{1+z}{4}\right)^6\,,
\label{meandepth}
\end{equation}  
where ${\cal A}(z)=\left(\om{b}h^2\right/0.02)^2 H_{100}(z)^{-1}
\Gamma_{-12}(z)^{-1}$. Here, $\om{b}$ is the baryon content and $H(z)$
the Hubble constant in unit of 100$\kmsmpc$ (e.g. Rauch \etal 1997;
McDonald \etal 2000).  To calculate $\Gamma_{-12}$, we assume
$\om{b}h^2=0.02$, in agreement with the constraint derived from CMB
and deuterium abundance  measurements (e.g. Spergel \etal 2003;
Burles \& Tytler 1998).  For the models of Fig.~\ref{fig5} which best
fit the PS+PDF data, the  optical depth normalisation is
$\tau_0\sim$0.9 (scale-invariant) and  $\sim$0.8 (RSI). The inferred
photoionisation rate for both best fit models  is $\Gamma_{-12}\sim
0.6$, a value consistent with other estimates  (e.g. Rauch \etal 1997;
McDonald \& Miralda-Escud\'e 2001). Note, however, that this does not
ensure that our simulations fully resolve the \op forest 
(cf. Section~\S\ref{spectra}).

\subsection{Sensitivity to the clustering amplitude}
\label{sigma8}

As we discussed before, a scale-invariant model with a clustering
amplitude of $\sigma_8=0.9$ matches best the data when the IGM
temperature is low.  Nonetheless, we could improve the agreement with
the data in the range $\thgg\bsim 1.5$  by decreasing the rms density
fluctuations.  To assess the sensitivity of the statistics to
$\sigma_8$, we take advantage of the self-similarity of gravitational
clustering in EdS cosmology (a good approximation at redshift $z\bsim
2$).  As a result, a snapshot at redshift different from $z=3$, once
rescaled,  can mimic a cosmology with a different $\sigma_8$.  Since
the fluctuation amplitude of the RSI model is tightly constrained by
the CMB and \op forest data (e.g. Bennett \etal 2003; Spergel \etal
2003), we will vary $\sigma_8$ in the scale-invariant cosmology only.
We consider snapshots separated by a redshift interval $\Delta z=0.2$
in  the range $z=2-4$. These snapshots thus mimic scale-invariant
cosmologies of rms density fluctuation $0.7\lsim\sigma_8\lsim 1.1$.

\subsubsection{Correlation between $\sigma_8$ and $\thg$}

\begin{figure}
\resizebox{0.45\textwidth}{!}{\includegraphics{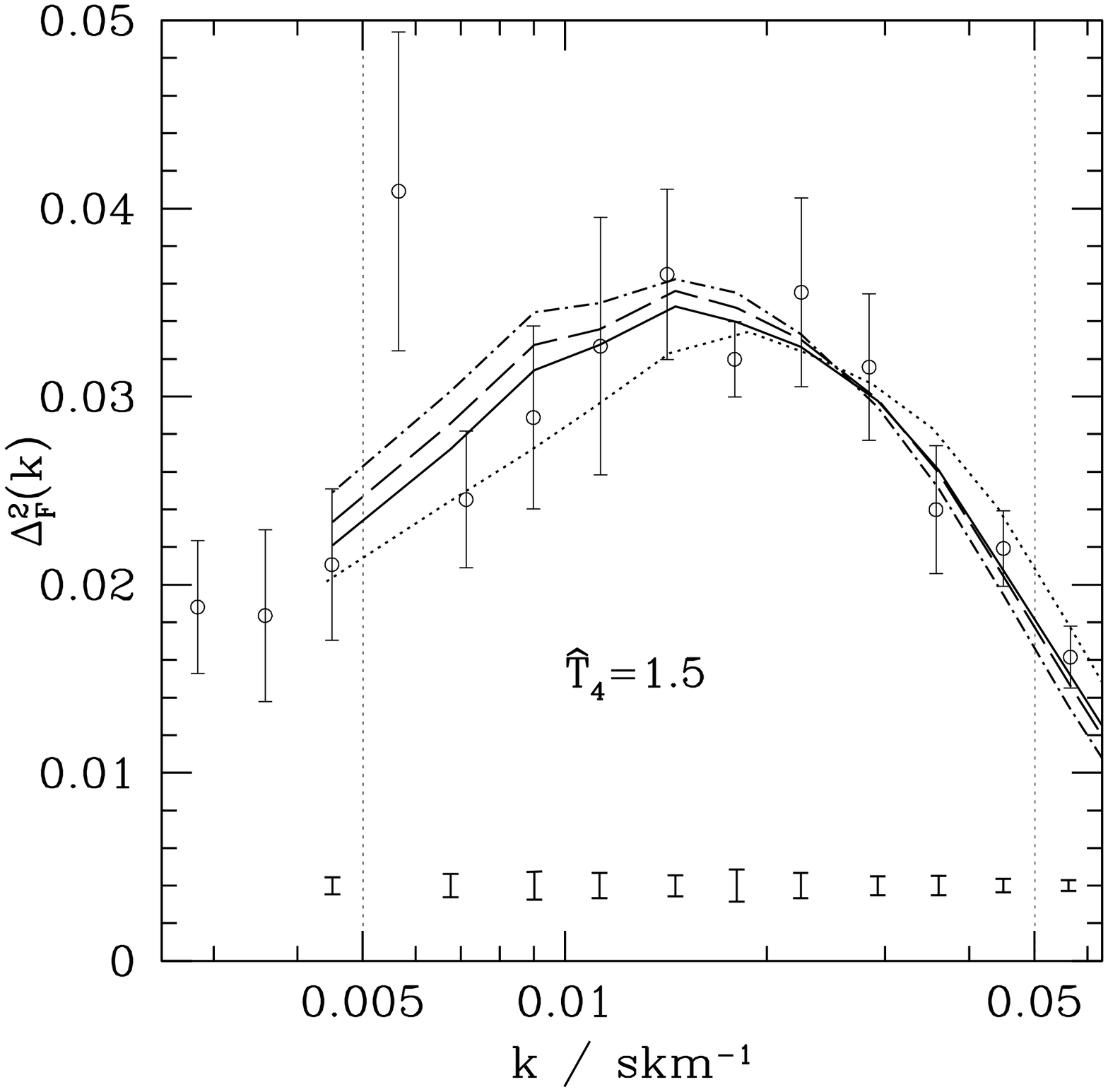}}
\resizebox{0.45\textwidth}{!}{\includegraphics{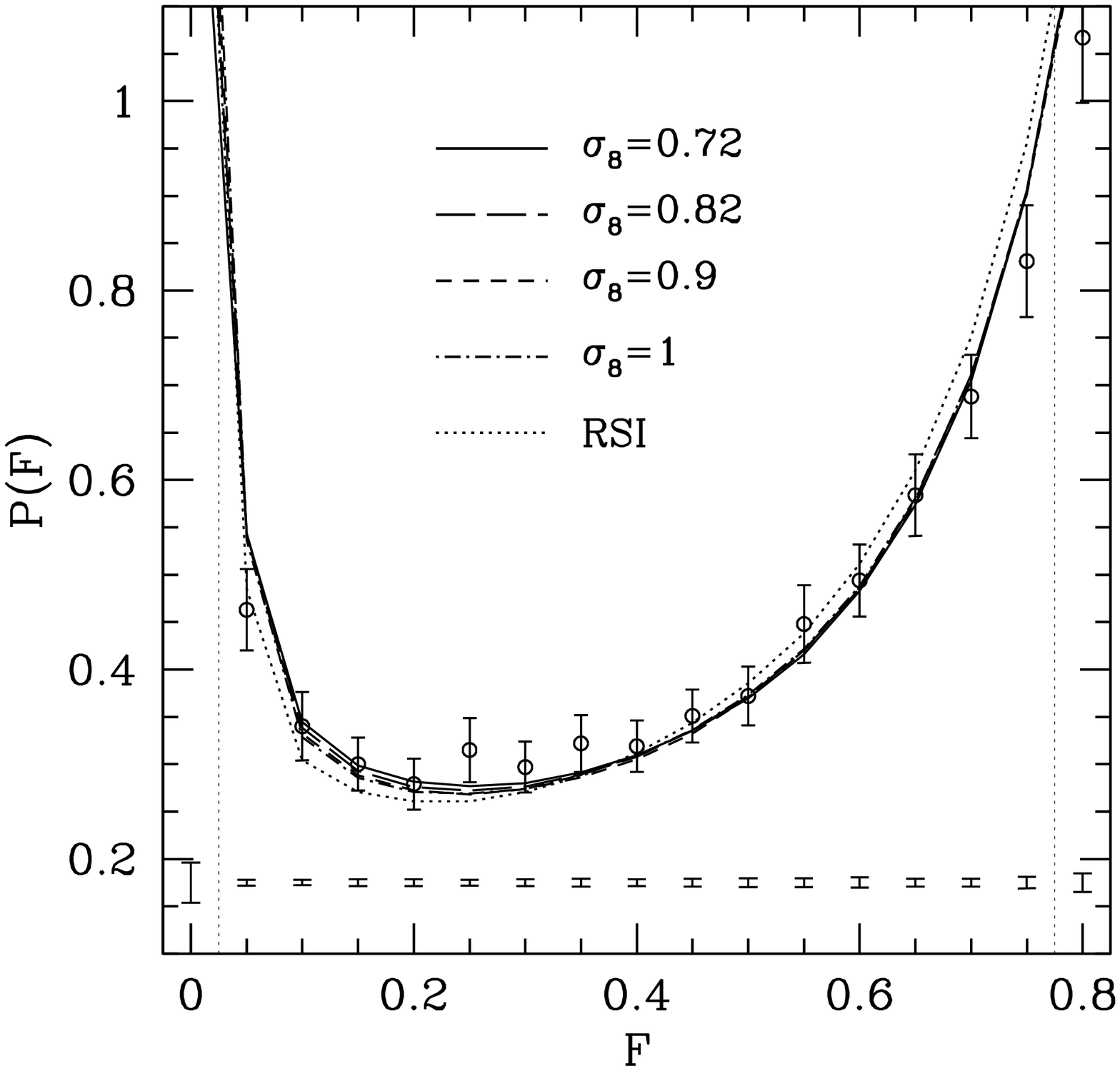}}
\caption{A comparison between the flux PS (top panel) and PDF (bottom
panel)  of scale-invariant models with rms density fluctuation
$\sigma_8=0.72$ (solid), 0.82 (long-dashed),  0.9 (short dashed) and 1
(dashed-dotted curve). The  RSI model is shown as dotted curves. The
IGM temperature has a fixed value $\thgg=1.5$, while the other
parameters  are varied to obtain the best agreement with the data. The
best fit models are listed in Table~\ref{table2}. In the bottom of
each panel,  the errorbars show our estimate of the cosmic variance
for the model with $\sigma_8=0.9$ (cf. text).}
\label{fig7}
\end{figure}

In Fig.~\ref{fig7},  we compare the PS (top panel) and the PDF (bottom
panel) of several best fit  models with different clustering
amplitudes~: $\sigma_8=0.72$ (solid),  0.82 (long-dashed), 0.9 (short
dashed) and 1 (dashed-dotted curves).  We also show the best fit RSI
model as a dotted curve. The mean IGM  temperature was chosen to be
$\thgg=1.5$. For sake of completeness however,  we list in
Table~\ref{table2} the parameters values of best fit models  for a
mean IGM temperature in the range $1\leq\thg\leq 2.5$.  A comparison
between the scale-invariant $\sigma_8=0.82$  model and the RSI which
has a very similar $\sigma_{8}$, demonstrates  that the best fit power
spectrum on scale $k\lsim 0.01\skm$ is very sensitive to the
small-scale behaviour of the matter power spectrum.  Indeed, the
difference in the flux power spectrum of these two models  mainly
results from including the PDF in the chi-squared statistics. This
illustrates the advantage  of combining several statistics of the \op
forest in constraining  the shape of the linear power spectrum.

The scale-invariant models  fit the flux power spectrum and the PDF
reasonably well. However, the large-scale amplitude of their flux PS
increases with $\sigma_8$, and causes the models with  $\sigma_8\bsim
0.9$ to overestimate the data on scale $k\lsim 0.01\skm$.  To assess
the significance of this effect, we need an estimate for the cosmic
variance error in the flux PS.  To this purpose, we ran two additional
simulations (S3 and S4) of a  scale-invariant cosmology with
normalisation $\sigma_8=0.9$ and computed  the flux PS and PDF for the
best fit models of Fig.~\ref{fig7}. Then, from  the sample of
simulations $S1\rarrow S4$, we computed the flux PS and PDF  for the
six possible combinations of two simulations. The cosmic variance
error was then the 1$\sigma$ scatter around the mean.  For clarity, we
use an offset and plot in Fig.~\ref{fig7} the cosmic  variance as
errorbars in the bottom of each panel.  The errorbars are shown for
the model with $\sigma_8=0.9$ only, as we found that they do not
change much among  the models. As we can see, they  are small, about
$\lsim$3\% and $\lsim$2\% in the  flux PS and PDF
respectively. However, since our simulations lack of  large-scale
power, they most probably underestimate  the true cosmic variance.
Yet the flux PS of the models with $\sigma_8\bsim 0.9$ would still
confidently lie above the data  points on scale $k\lsim 0.01\skm$,
even if the errors were twice as  large. This demonstrates that, at
fixed temperature, $\dfk$ increases with $\sigma_8$ on scale $k\lsim
0.01\skm$. Consequently, there is a  degeneracy between the
temperature and $\sigma_{8}$. A  model with larger $\sigma_8$ requires
a lower temperature to match the  observed flux power spectrum. Note,
however, that the measurement errors  are large in that range of
wavenumber. As a result, although the models  with $\sigma_8\bsim 0.9$
shown in Fig.~\ref{fig7}  predict a flux power spectrum which, ``by
eye'', overestimates the observations, they are still consistent with
the data in a $\chi^2$ sense at least (The worst chi-squared,
$\chi^2=25.4$ for 23 degrees of freedom, is obtained for $\sigma_8=1$,
and corresponds to a fit probability $\sim$30\%).

\begin{figure*}
\resizebox{0.9\textwidth}{!}{\includegraphics{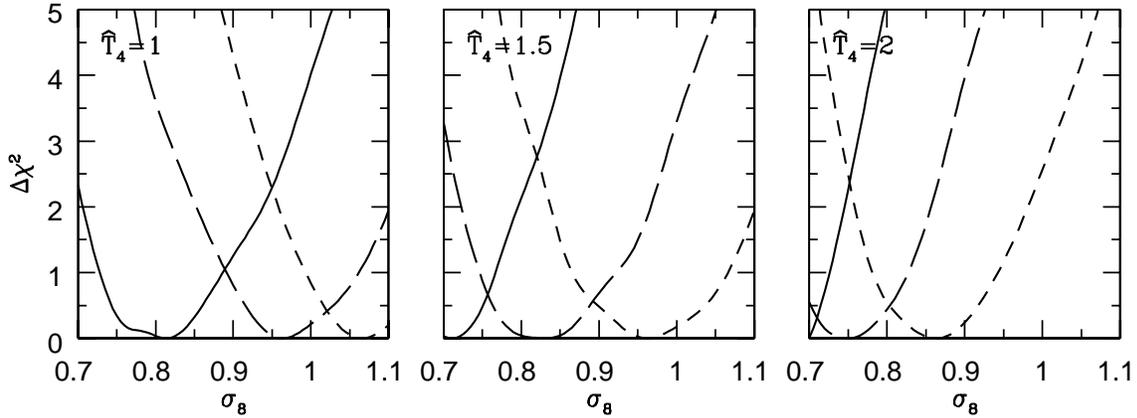}}
\caption{Constraint on $\sigma_8$ from a fit to the observed flux  PS
and PDF after marginalization over the the mean transmitted flux. Only
the parameters $\sigma_8$ and $\la F\ra$ were varied.  The curves show
$\Delta\chi^2=\chi^2-\chi^2_{\rm min}$ for a mean IGM  temperature
$\thgg=1$ (left panel), 1.5 (middle panel) and 2 (right  panel). The
filtering is $\kf=8.3$ (solid curve), 7.1 (long-dashed)  and
$6.3\hmmpc$ (short-dashed). The adiabatic index has a fixed value
$\gamma=1.3$.}
\label{fig8}
\end{figure*}

To illustrate the correlation between the mean IGM temperature  and
$\sigma_{8}$, we plot in Fig.~\ref{fig8} the difference
$\Delta\chi^2= \chi^2-\chi_{\rm min}^2$ as a function of $\sigma_8$,
after marginalizing over the mean flux level $\la F\ra$. The other
parameters do not vary in the chi-squared minimization.  The mean IGM
temperature is $\thg=1$  (left panel), 1.5 (middle panel)  and 2
(right panel), the filtering is $\kf=$8.3 (solid curve), 7.1
(long-dashed) and  6.3$\hmmpc$ (short-dashed), and the adiabatic index
has a fixed value $\gamma=1.3$.  These values of $\thg$ and $\gamma$
are consistent with that inferred from  observations at $z\sim 3$
(e.g. Schaye \etal 2000; Ricotti, Gnedin \& Shull  2000; McDonald
\etal 2001).   The reason for selecting particular values of $\gamma$
and $\kf$ follows from the fact that marginalizing over these
parameters proves difficult because of their degeneracy. The large
measurement errors worsen the situation.  However, we can take
advantage of the fact that the subspace $\kf$-$\gamma$ has one
effective degree of freedom. Namely, we can fix the adiabatic index
$\gamma$ and select several values of the filtering length $\kf$
without restricting the analysis.  Note that the wiggles which appear
in $\Delta\chi^2$ are caused by numerical instabilities in the spline
interpolation.  Fig.~\ref{fig8} demonstrates that, at a given
filtering $\kf$,  decreasing the temperature  increases the
normalisation $\sigma_8$ of the best fit models
($\Delta\chi^2=0$). Moreover, at fixed temperature, decreasing the
filtering $\kf$ also increases the best fit $\sigma_8$. For e.g.
$\thgg=1.5$, we find  $\sigma_8\approx 0.71$, 0.83 and 0.96 for
$\kf$=8.3, 7.1 and  6.3$\hmmpc$. In other words, one  has to increase
the clustering amplitude in order to compensate for the larger
smoothing.  It would therefore be possible to match the data with  a
normalisation $\sigma_8\bsim 1$ if both the filtering and the IGM
temperature were very low, $\kf\approx 6\hmmpc$ and $\thgg\approx 1$.
Indeed, a model where $\kf=6.3\hmmpc$ and $\thgg=1$ has a best fit
value  $\sigma_8=1.06$, and fits the data with an acceptable
chi-squared   $\chi^2=22.4$ for 24 degrees of freedom.

\subsubsection{A temperature-dependent filtering}
\label{tdep}

As we discussed in the previous Section, a model with $\sigma_8\bsim
1$ can match the data if both $\thg$ and $\kf$ are very low. In
reality, however, these two parameters are not independent since 
the filtering $\kf$ to increase with the temperature (when all
the other parameters are fixed). To make further progress, we will
assume that the filtering $\kf$ is solely a function of the mean IGM
temperature $\thg$. In Appendix~\S\ref{filtering}, the dependence of
$\kf$ on the IGM temperature  and the reionization history  of the
Universe is discussed in detail. In brief, we expect the filtering
scale to decrease with increasing temperature as $\kf\propto
\thgg^{-1/2}$, and to be smaller than $\sim 14\hmmpc$ at $z=3$ for
reasonable  reionization scenarios (cf. Appendix~\S\ref{filtering}).

\begin{figure*}
\resizebox{0.40\textwidth}{!}{\includegraphics{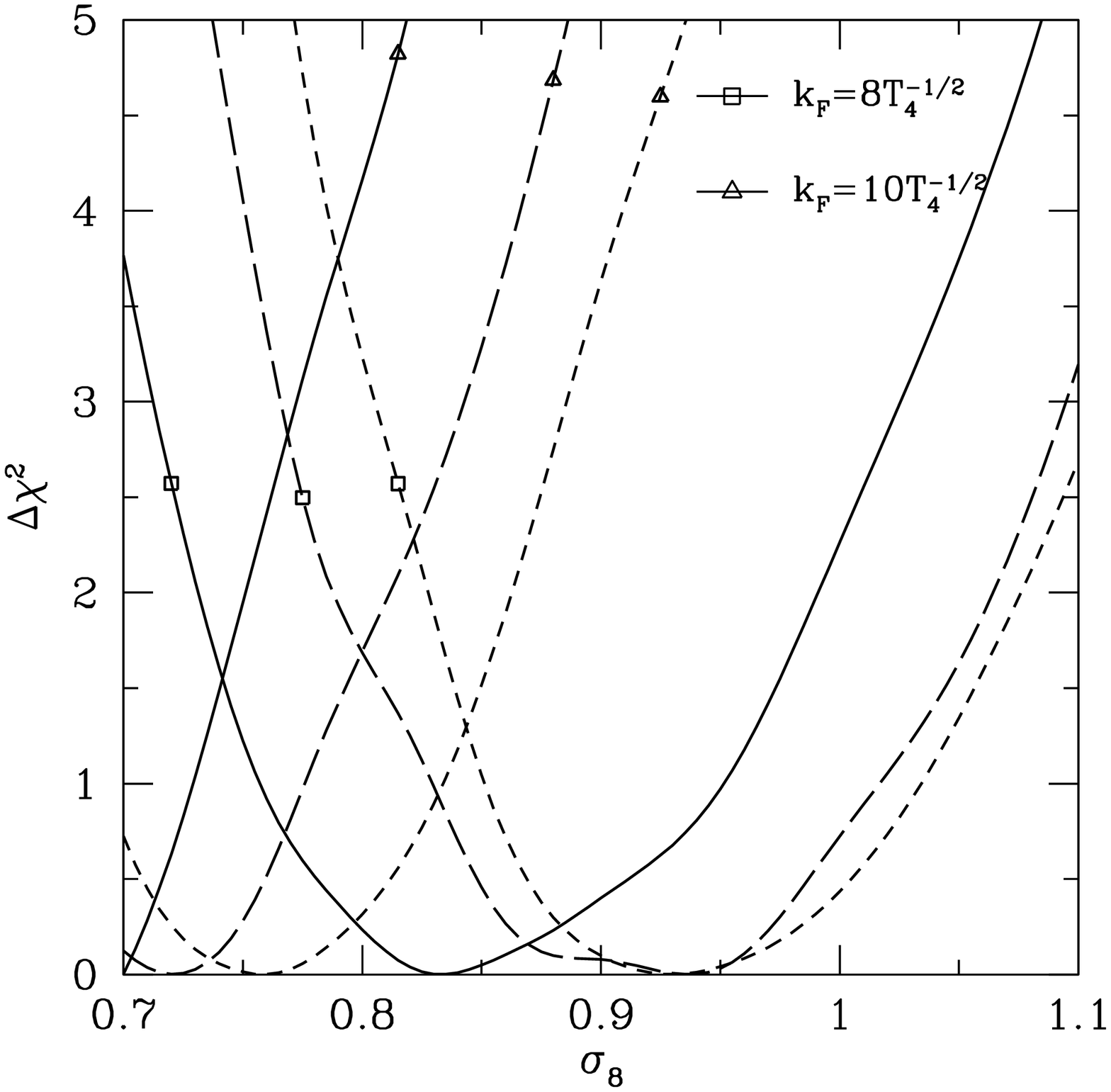}}
\resizebox{0.52\textwidth}{!}{\includegraphics{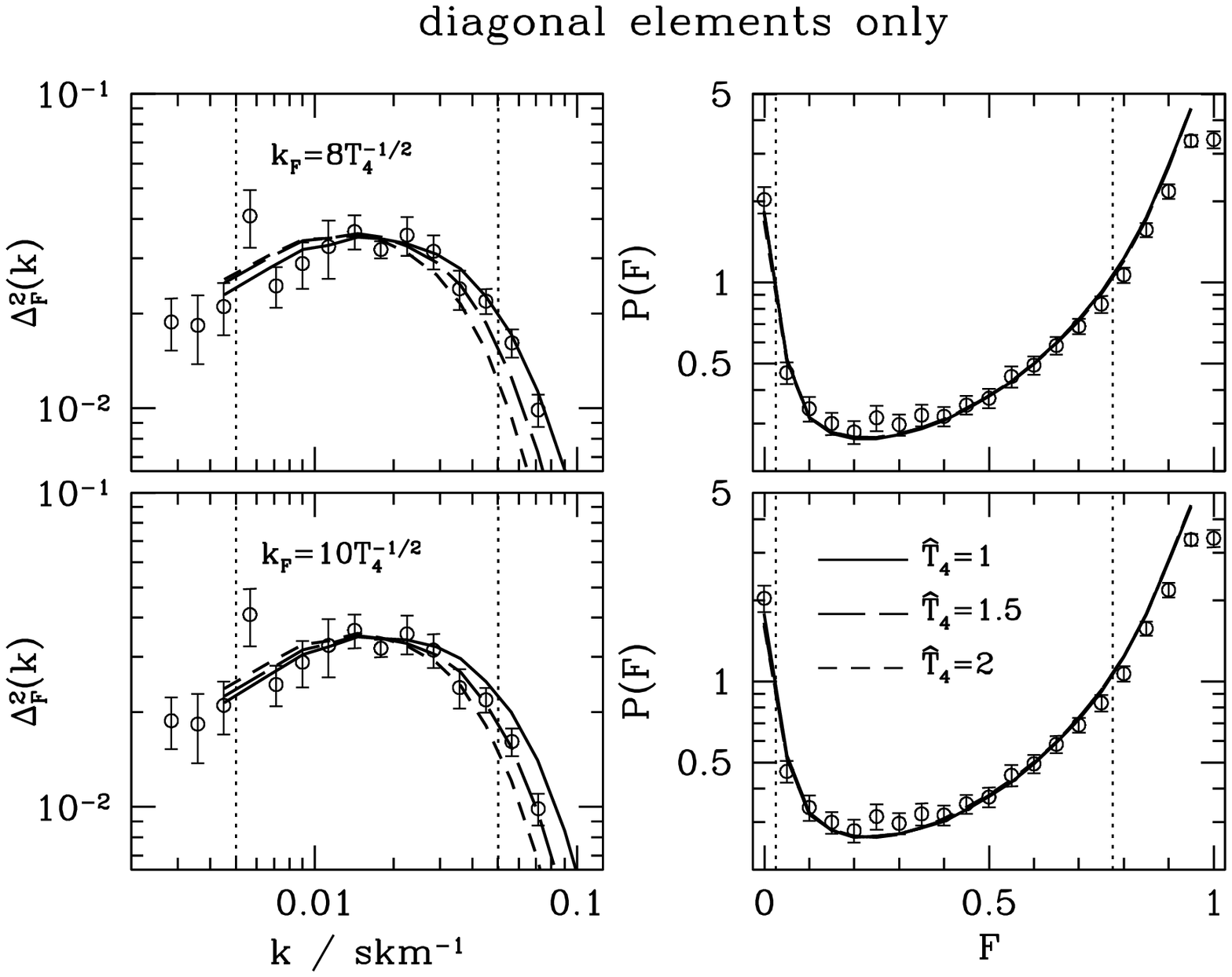}}
\caption{{\it Left panel}~: $\Delta\chi^2$ after marginalizing over
the mean flux level. Only $\sigma_8$ and $\la F\ra$ were varied to
obtain the best fit parameters. The mean IGM temperature $\thgg=1$ (solid
curves), 1.5 (long-dashed) and 2 (short-dashed). The filtering scales
with $\thgg$ as $\kf=8\,\thgg^{-1/2}$ (curves with squares) and
$10\,\thgg^{-1/2}\hmmpc$ (curves with triangles). The adiabatic index
has a fixed value $\gamma=1.3$. The spline interpolation was
performed using  the diagonal elements of the PDF covariance matrix
solely.  {\it Right panel}~: The flux power spectrum and PDF of the
best fit   models ($\Delta\chi^2=0$). The IGM temperature is $\thgg=1$
(solid curves), 1.5 (long-dashed) and 2 (short-dashed), for a fixed
adiabatic index $\gamma=1.3$.  The best fit values of $\sigma_8$, $\la
F\ra$  together with the chi-squared are listed in the upper half of
Table~\ref{table3}.}
\label{fig9}
\end{figure*}

To assess the effect of such an assumption on the constraint inferred
from the flux PS and PDF, we plot  in the left panel of
Fig.~\ref{fig9} $\Delta\chi^2$ as  a function of $\sigma_8$ after
marginalizing over the mean flux. Only $\sigma_8$ and $\la F\ra$ were
varied to obtain the flux PS and PDF of the best fit models shown in
the right panel. We assumed a temperature-dependent  filtering scaling
as $\kf=10\thgg^{-1/2}$ (curves with  triangles) and
$8\thgg^{-1/2}\hmmpc$ (curves with squares). The  results are shown
for different values of the temperature, $\thgg=1$ (solid curves), 1.5
(long-dashed) and 2 (short-dashed), but for a fixed value of the
adiabatic index, $\gamma=1.3$. The parameter values of the best fit
models are listed in the upper half of Table~\ref{table3}.  The
temperature-dependent filtering causes the best fit value of
$\sigma_8$ to increase with the temperature. The rms fluctuations even
reaches  $0.9-0.95$ for  a filtering $\kf=10\thgg^{-1/2}$ and a
temperature $\thgg\lsim 1.5$. Most importantly, the amount of
filtering significantly matters, as the  best fit parameters are
rather sensitive to the normalisation of the relation $\kf(\thgg)$. At
fixed temperature, a scaling $\kf=8\thgg^{-1/2}$ leads to best fit
values of $\sigma_8$ which are 10-20\% larger than a scaling
$\kf=10\thgg^{-1/2}$.  This strongly suggests that the Gaussian filter
approximation (e.g. Gnedin \etal 2003) is certainly not accurate
enough to constrain the clustering amplitude to better than
$\sim$10\%.  A close look at the right panel of Fig.~\ref{fig9} also
reveals that the  models with $\kf(1)=10\hmmpc$~\footnote{$\kf(1)$
stands for $\kf(\thgg=1)$.} look consistent with the data,   whereas
those with $\kf(1)=8\hmmpc$ tend to overestimate $\dfk$ on scale
$k\lsim 0.01\skm$, especially when  the temperature is large,
$\thgg\bsim 1.5$. The best fit values of the clustering amplitude
inferred from a temperature-dependent filtering $\kf\leq
10\thg^{-1/2}$ are $\sigma\bsim 0.7$. However, since the Nyquist
frequency of the grid is  $k_{\rm Ny}\approx 10.2\hmmpc$, we cannot
robustly probe the range of  wavenumber  $\kf(1)\bsim 10\hmmpc$.
Consequently, our analysis cannot exclude values of $\sigma_{8}$ as
low as $\sigma_8\leq 0.7$.  However, since $\kf(1)=14\hmmpc$ is probably
an unrealistically high value for the filtering wavenumber at
redshift $z=3$ (Appendix~\S\ref{filtering}), we expect
$0.7\lsim\sigma_8\lsim 0.9$ from the present results for a  reasonable
choice of reionization history, and an IGM temperature in the range
$1\lsim\thgg\lsim 2$.

\subsubsection{Effect of including the full PDF error matrix}
\label{matrix}

\begin{figure*}
\resizebox{0.40\textwidth}{!}{\includegraphics{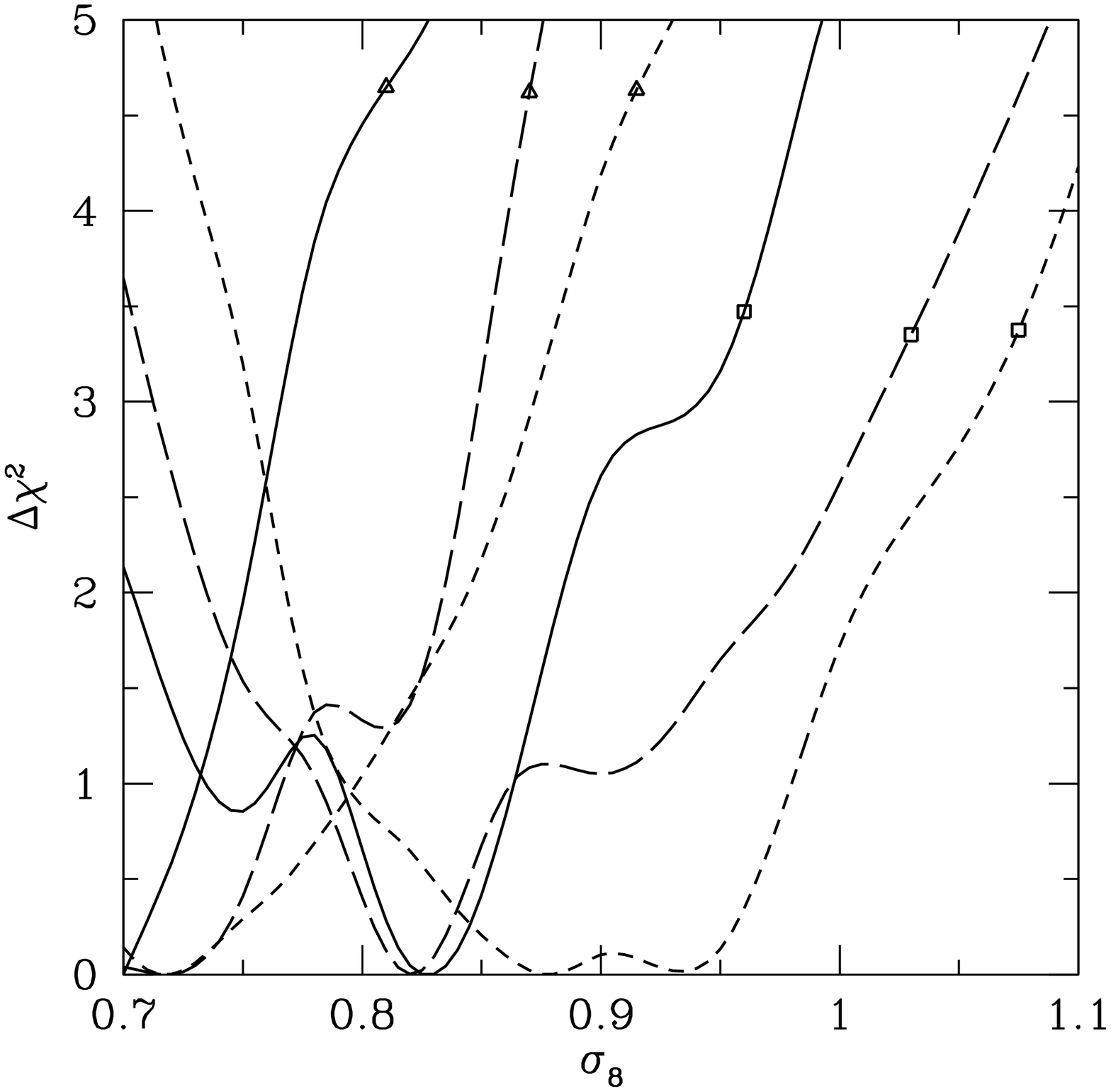}}
\resizebox{0.52\textwidth}{!}{\includegraphics{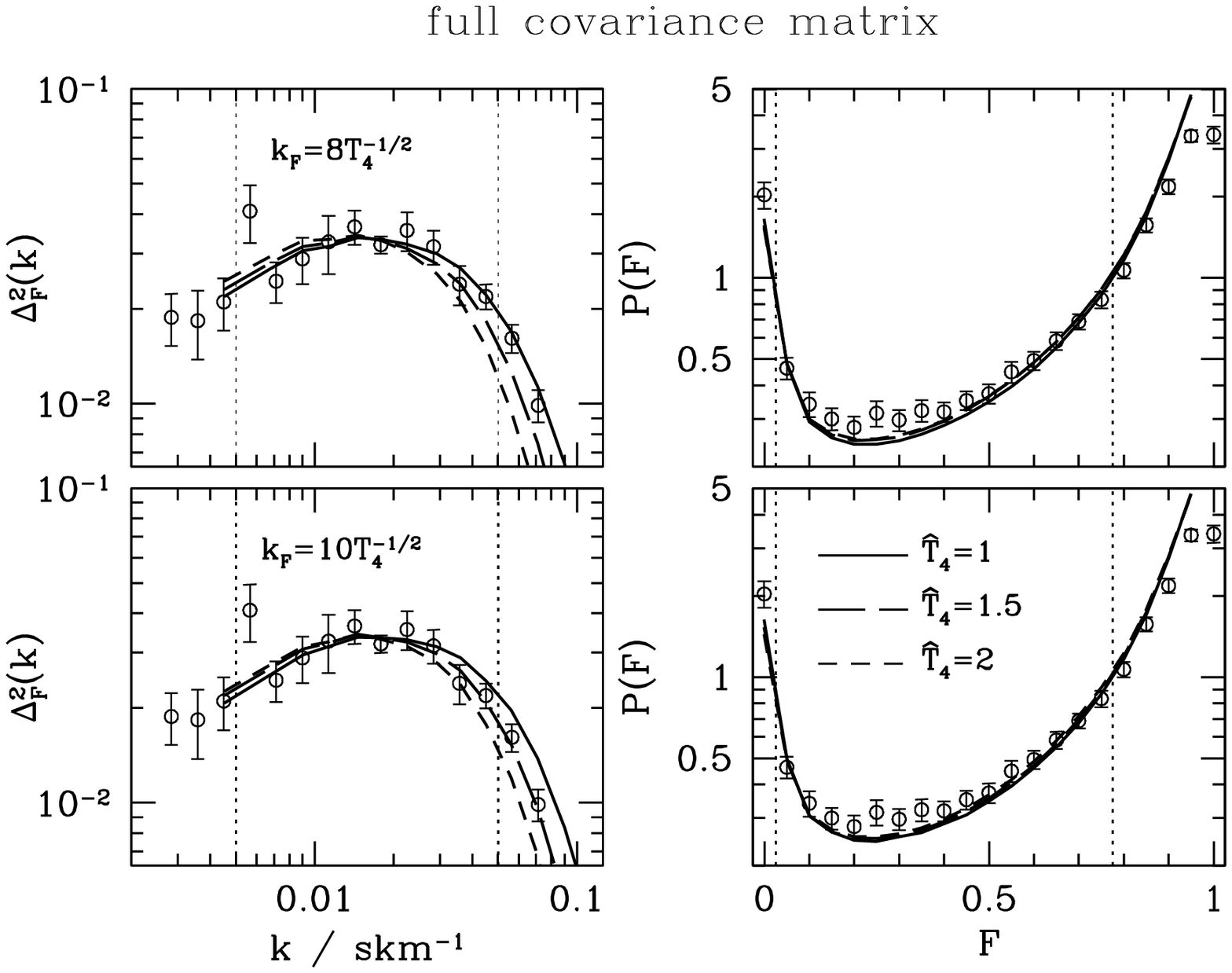}}
\caption{Same as Fig.~\ref{fig9}. However, the full covariance matrix 
of the flux PDF (cf. text) was used instead of the diagonal elements 
only. The best fit parameter values are listed in the lower half of 
Table~\ref{table3}.}
\label{fig10}
\end{figure*}

As yet we have  set to zero all the off-diagonal elements of the
error matrix. We now present results with the full error covariance
matrix as given in McDonald \etal (2001). A comparison between the
results in the two cases allows an assessment of the robustness of the
inferred best fit parameters.  Since we are interested in the
measurements with $0.05\leq F\leq 0.75$, we obtain the desired
covariance matrix by setting the errors of the data points outside
that range to infinity.  Curves of $\Delta\chi^2$ versus $\sigma_{8}$
are plotted  in  the left  of Fig.~\ref{fig10}.  In the panel to the
right of the same figure flux power spectra and PDFs corresponding to
the best fit parameters are shown.  Values of the best fit parameters
are listed in the lower half  of Table~\ref{table3}. Taking into
account the covariance matrix of the flux PDF in the chi-squared
minimization introduces a systematic shift in  the central values of
the best fit parameters.  Yet the effect is rather small. The
clustering amplitude, $\sigma_8$,  gets lower by only a few percents
in most  cases.    The wiggles in the chi-squared curves in the left
panel of Fig.~\ref{fig10} lead to a highly non-gaussian  distribution
for the statistical  errors on the best fit parameters.   As a result,
assessing to which extent the errors on the best fit values  of
$\sigma_8$ compare with those inferred from the diagonal elements only
proves difficult.  However, we found that at fixed $\sigma_8$ and
$\thg$, including the  off-diagonal elements increases the confidence
levels in the plane $\kf$ - $\gamma$ by 5-10\%
(cf. Section~\S\ref{plane}).  Finally, a comparison between the right
panels of Fig.~\ref{fig9}  and \ref{fig10} reveals that, when the full
covariance matrix is included, the flux PDF of the best fit models
falls systematically below the data points in the range $0\lsim F\lsim
0.5$. This kind of behaviour is likely to occur when the errors are
strongly correlated,  as it is the case for the flux PDF. The curve
for which the likelihood is  the largest is not necessarily the one
which goes through the error bars attached  to the data points.

\subsection{The observed line statistics}
\label{line}

To assess whether the best fit models reproduce the observed line
statistics,  we compute the column density and line width
distributions, and compare our  results to observational data.

\subsubsection{The column density distribution}

The differential line density distribution $f(\nhi)$ is defined as the
number  of lines per unit column density and per unit absorption
distance (see e.g. Tytler 1987; Schaye 2001), which is  $\dd X/\dd
z\approx \om{m}^{-1/2}(1+z)^{1/2}$ at redshift $z\sim 3$ in the
standard $\Lambda$CDM model. However, to facilitate the comparison
with the measurements of Petitjean \etal (1993) and Hu \etal (1995),
who  assumed a deceleration parameter $q_0=0$, we take the absorption
distance to be  $\dd X/\dd z=(1+z)$. In the left  panel of
Fig.~\ref{fig11} we plot the column  density distribution for several
models which give a good fit of the PS and PDF  data ($\chi^2/\nu\lsim
1$). $f(\nhi)$ was calculated from a sample of 40  LOS.  The simulated
curves are compared to the data of Hu  \etal (1995) and Petitjean
\etal (1993), which are shown as filled  circles and squares
respectively. The mean redshift of both samples is $z=2.8$, close  to
that of the M00 data considered in the present paper. We plot
$f(\nhi)$ for two scale-invariant, $\sigma_8=0.9$ models with
different values of $\kf$ and $\gamma$~: $(\kf,\gamma)=(8.3,1.5)$
(solid curve) and $(6.3,1.1)$ (solid-dotted curve). In addition, we
show $f(\nhi)$ for a cold ($\thgg=1.5$) and a hot ($\thgg=2.5$)
scale-invariant model of normalisation $\sigma_8=0.72$ as long- and
short-dashed curves respectively. The RSI model is shown as a dotted
curve.  Absorption lines with column density  $10^{12.5}\leq \nhi\leq
10^{15.5}\cmm$ contribute most to the \op forest.  In that range, the
simulated column density distributions agree very well  with the data.
In particular, the slope and the normalisation of the simulated
distributions coincides with that of the observations. The later also
tends to steepen in the range $\nhi\bsim 10^{14}\cmm$. Such a
deviation  from a single power law is also found in the observations
(e.g. Petitjean \etal 1993; Kim \etal 1997).  Fig.~\ref{fig11}
demonstrates that, although the flux PS and PDF of the best  fit
models is rather sensitive to the normalisation amplitude and the
relative amount of small-scale power (cf. Section~\S\ref{sigma8}), the
column density distribution barely changes among the various models,
in agreement with the results of Theuns, Schaye \& Haehnelt (2000) who
found that $f(\nhi)$ is insensitive to the amount of small scale
power. If most of the lines identified by VPFIT have a column density
$\nhi\lsim 10^{15.5}\cmm$ ($\sim 10^3$ for a sample of 40 LOS),
absorption systems with column density exceeding  $\nhi\bsim
10^{15.5}$ are rare ($\lsim 30$). It is however crucial to reproduce
them since they can contribute up to 50\% of the flux  power spectrum
on scale $k\lsim 0.01\skm$ (e.g. Viel \etal 2004b).  Although the
column density distribution suffers from the shot noise associated to
these rare events, the left panel of Fig.~\ref{fig11}  shows clearly
that our simulations reproduce, at least qualitatively,  these strong
absorption systems.

\subsubsection{The line-width distribution}

\begin{figure*}
\resizebox{0.45\textwidth}{!}{\includegraphics{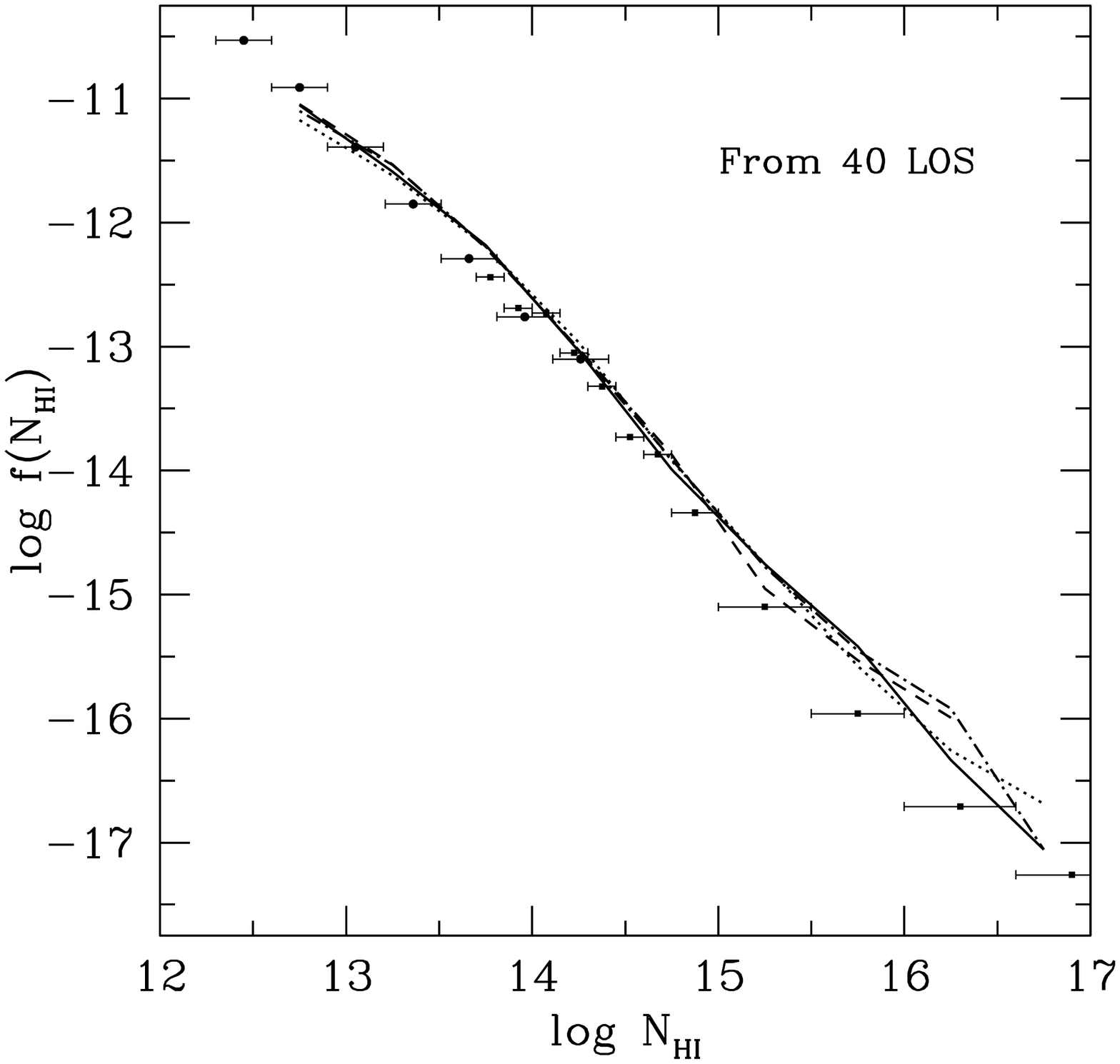}}
\resizebox{0.45\textwidth}{!}{\includegraphics{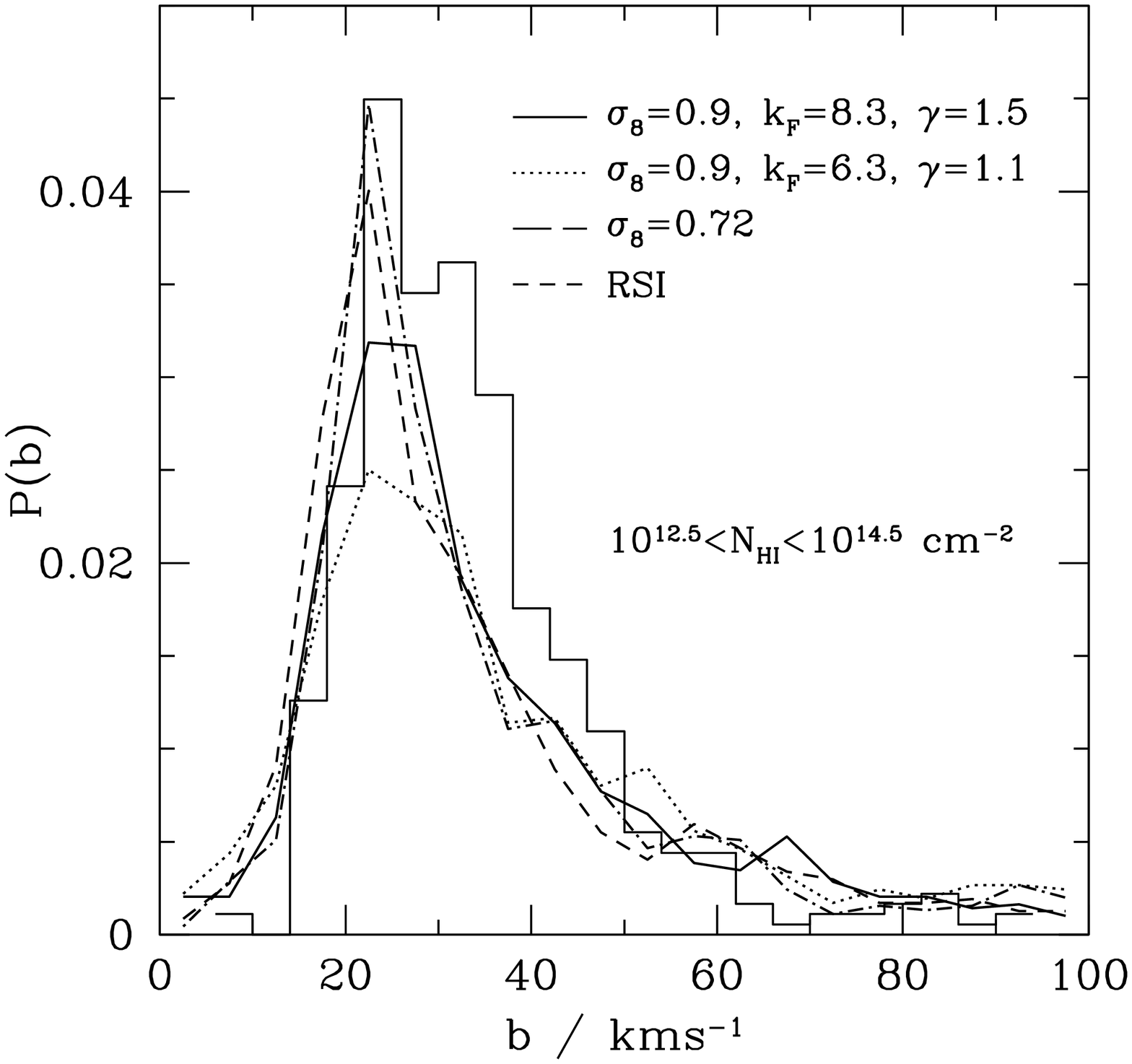}}
\caption{{\it Left panel}~: The differential column density
distribution  $f(\nhi)$ of various models which fit the PS and PDF
data best~: two scale-invariant models of normalisation $\sigma_8=0.9$
(solid and dotted-dashed curve), a scale-invariant model with
$\sigma_8=0.72$ (long-dashed curve),  and a RSI model (short-dashed
curve). The parameter values of these models are  $(\kf,\la
F\ra,\gamma,\thgg)$=(8.3,0.69,1.5,1.5),  (6.3,0.69,1.1,1.5),
(10,0.7,1.5,1.5) and (12.5,0.7,1.1,2)  respectively. Note that the
scale-invariant models have a temperature $\thgg=1.5$, while the RSI
model has $\thgg=2$.  The filled circles and squares are the data of 
Hu \etal (1995) and  Petitjean \etal (1993) respectively.  {\it Right
panel}~: the corresponding line-width distribution $f(b)$.  Only the
lines with column density $\nhi=10^{12.5}-10^{14.5}\cmm$ were  taken
into account. The simulated distributions are compared to the data of
Hu \etal (1995) shown as an histogram.}
\label{fig11}
\end{figure*}

The line-width distribution, $f(b)$, which is the fraction of line
with  a given width, is shown in the right panel of Fig.~\ref{fig11}
for the models plotted in the left panel. $f(b)$ is computed from
absorption systems with column density $10^{12.5}-10^{14.5}\cmm$. The
solid histogram is  the data of Hu \etal (1995). The observed
line-width distribution  clearly exhibits a peak in the range $b\sim
20-40\kms$. In this  respect, Fig.~\ref{fig11} shows that in the
simulations, the peak is  more pronounced for a model with larger
value of $\gamma$ at constant $\sigma_8$. Furthermore, the
scale-invariant model with lower normalisation ($\sigma_8=0.72$) and
the RSI model account somewhat better  for the amplitude of the
observed peak, $f(b)\sim 0.04$.  However, given the large measurement
errors, using $f(b)$ to constrain the parameter of the model does 
not seem feasible. Most importantly, it has been demonstrated 
that very high-resolution hydro-simulations are needed to reliably 
predict $f(b)$ (Theuns \etal 1998; Bryan \etal 1999; Theuns, Schaye 
\& Haehnelt 2000). Consequently, it is difficult to draw any firm 
conclusion on the sensitivity of the line-width distribution to the 
shape and amplitude of the matter power spectrum of the best fit 
models.

\subsection{Systematic errors}
\label{errors}

Various sources of error affect our analysis, among them systematics
in the data and in the modelling of the \op forest.  Regarding the
data, note that the M00 measurements, obtained from a sample of 8 QSO
only, are quite noisy. In the flux power spectrum for example, the
data point at $k=0.00566\skm$ lies well above the others. One might
be worried that our results are strongly affected by this single data
point. However,  the corresponding error, $\sim$20\%, is larger than
that of  the other data points. For  the best fit model with
normalisation $\sigma_8=0.9$ shown in Fig.~\ref{fig7}, the
contribution  of this data point to the chi-squared of the PS
measurements  ($\chi^2=10.4$) is $\Delta\chi^2=2.86$, smaller than
that of the data  point at $k=0.00713\skm$, for which
$\Delta\chi^2=4.45$. We found indeed that the removal of this data 
point from the chi-squared does not  affect the results noticeably.

Continuum fitting errors are likely to affect the flux probability
distribution. The modelling of these errors is complicated by the fact
that the scales of interest are of the order of the box size $L$ of
the simulations.  However, M00 were able to demonstrate that, if the
inclusion of continuum fitting errors can account for most of the
discrepancy between the simulated and observed PDF in the range
$F\bsim 0.8$, it should not affect much the PDF for $F\lsim 0.8$. We
thus believe that the PDF in the range $F\lsim 0.8$ is robust to these
errors.  Continuum fitting errors might also affect the inferred
column densities $\nhi$. Notwithstanding, the deviation  from a single
power-law in the regime $\nhi\lsim 10^{14}\cmm$ is most  probably
caused by line blending, and by the nonlinear evolution of the
structures associated to the absorption systems (e.g. Kim \etal 2002;
see also Schaye 2001).  Finally, continuum fitting errors should play
a minor role in the flux power spectrum as the scales probed by our
simulations are relatively small,  $k\bsim 0.003\skm$ (e.g. Kim \etal
2004).

Up to now we have adopted the prior  $\la F\ra=0.684\pm
0.023$ (M00). However, one should bear in mind that the presence of
metal lines or strong absorption systems can substantially affect the
determination of $\la F\ra$ (e.g. Schaye \etal 2003; Viel \etal
2004b).  In particular, Schaye \etal (2003) found that accounting for
strong  \hi\ systems and metal lines   
yields a lower value of $\la F\ra=0.637$ at $z=3$.  Furthermore, as
shown by Seljak, McDonald \& Makarov (2003) and Viel, Haehnelt \&
Springel (2004), a wrong estimate of the mean flux level can severely
bias the results.  It is therefore prudent to examine to which extent
the assumed mean  flux level affects the results. For this purpose, we have
repeated the  analysis of Section~\S\ref{tdep} with a mean flux $\la
F\ra=0.637$ (Schaye \etal 2003), and a standard deviation
$\sigma_{\bar{F}}=0.030$. Only the diagonal elements of the PDF error
matrix were considered in the  chi-squared minimization. Furthermore,
$\Delta\chi^2$ was calculated  from the PS data alone, and from the
whole PS+PDF data set. We found that, in both cases, lowering the mean
flux level amounts to a larger best fit value of $\sigma_8$, in
agreement with previous studies  (e.g. Viel, Haehnelt \& Springel
2004). However, the effect is much  weaker when $\Delta\chi^2$ is
computed from the whole PS+PDF data set ($\lsim$3\%) than from the PS
data alone ($\lsim$8\%). Combining the flux PS measurements with other
statistics of the \op forest can thus significantly reduce the
sensitivity of the results to the mean transmitted flux.

Our semi-analytical model of the \op forest allows us to explore a
much larger parameter space than full  hydrodynamical
simulations. However, the model  has several shortcomings.  It
neglects any possible  scatter in the temperature-density relation of
the low density IGM  as a results of shocks and  patch helium
reionization (e.g. Gleser \etal 2005). It also assumes a  filtering
length that is independent of  the local gas density and
temperature. It also ignores any galactic feedback (e.g. Croft \etal
2002a; Kollmeier \etal 2003).  Hydrodynamical simulations predict that
shock heating should drive a significant fraction of the baryons into
the warm-hot phase of the intergalactic medium (WHIM) at low redshift.
At the present epoch, this fraction might be as large as 40\%
(e.g. Cen \& Ostriker 1999; Dav\'e \etal 2001; see also Nath \& Silk
2001). At redshift $z\sim 3$, however, this fraction falls below
10\%. Moreover, numerical simulations also show that most of the WHIM
baryons at that redshift resides  in overdensities $\dgnl\bsim 10$
(Dav\'e \etal 2001). Hence, shock heating should have a rather weak
impact on  the low density IGM at $z\approx 3$.  Inhomogeneities in
the UV background might also affect the flux power spectrum. Yet this
effect should not be too important on the scales ($k\bsim 0.05\skm$)
and redshifts ($z=3$) of interest (e.g. Croft 2004;  McDonald \etal
2004c).  Recent measurements of the \op absorption near Lyman-break
galaxies (Adelberger \etal 2003) are taken as evidence for the
existence of dilute and highly ionised gas bubbles caused by
supernovae-driven winds. Notwithstanding, simulations indicate that
their small filling factor results in a moderate impact on statistics
of the \op forest such as  the power spectrum or the PDF (e.g. Croft
\etal 2002a; Weinberg \etal  2003; Desjacques \etal 2004; McDonald
\etal 2004c). Gleser \etal (2005) have demonstrated that  patchy
helium II reionization can cause significant scatter in the
temperature density relation.  A detailed account of the effect of
this scatter on the estimation of cosmological parameters has not yet
been taking into account neither in semi-analytic modeling nor in
hydrodynamical simulations.  A constant filtering length  also   lead
to systematic errors  in the inferred values of $\sigma_8$ (e.g. Viel,
Haehnelt \& Springel  2004).  In any case all of these effects are not
expected to amount to more than an error of 20\% in the estimated
best fit parameters.

\section{discussion}
\label{discussion}

\begin{table}
\caption{Parameter values of the models which best fit the flux PS and
PDF, assuming a temperature-dependent smoothing
$\kf\thg^{1/2}=$const. The  values listed in the upper half of the
table were obtained using only the diagonal  terms of the PDF error
matrix whereas, in the lower half, they were obtained  using the full
covariance matrix of the flux PDF.  The filtering scale $\kf(1)$ is
given in unit of $\hmmpc$.  The adiabatic index has a fixed value
$\gamma=1.3$, while the mean  IGM temperature is $\thgg=1$, 1.5 and
2. The last columns gives the  best fit values of $\la F\ra$,
$\sigma_8$ and the corresponding  $\chi^2$ (for 24 degrees of
freedom). }
\vspace{1mm}
\begin{center}
\begin{tabular}{cccccc} \hline
filtering & $\gamma$ & $\thgg$ & $\la F\ra$ & $\sigma_8$ & $\chi^2$ \\ 
\hline
$\kf(1)=10$ & 1.3 & 1 & 0.68 & 0.67 & 24.9 \\
            &     & 1.5 & 0.69 & 0.72 & 21.7 \\
            &     & 2 & 0.69 & 0.76 & 25.3 \\
$\kf(1)=8$  & 1.3 & 1 & 0.68 & 0.84 & 22.4 \\
            &     & 1.5 & 0.68 & 0.94 & 25.4 \\
            &     & 2 & 0.68 & 0.93 & 38.7 \\
\hline
$\kf(1)=10$ & 1.3 & 1 & 0.70 & 0.70 & 26.9 \\
            &     & 1.5 & 0.71 & 0.72 & 23.9 \\
            &     & 2 & 0.71 & 0.72 & 27.7 \\
$\kf(1)=8$  & 1.3 & 1 & 0.71 & 0.83 & 23.8 \\
            &     & 1.5 & 0.71 & 0.82 & 27.3 \\
            &     & 2 & 0.71 & 0.88 & 38.1 \\
\hline\hline
\end{tabular}
\end{center}
\label{table3}
\end{table}

We have used N-body simulations of  variants of the $\Lambda$CDM
models to generate  synthetic spectra of the \op forest.  We have
simulated the standard $\Lambda$CDM cosmology with scale-invariant
spectral index models, as well as a running spectral index model
(Spergel \etal 2003).  The one-dimensional flux power spectrum (PS)
and flux probability  distribution function (PDF) have been computed
from mock catalogs, and compared to the  observational data in order
to constrain the cosmological models and the  physical parameters that
dictate the properties of the \op forest.  In addition to the one- and
two-point correlations, we have also computed the neutral hydrogen
column density distribution and the line width distribution.  These
last two statistics have only been used as a consistency check for the
models which give the best fit to  the observed PS and PDF.

\begin{figure}
\resizebox{0.45\textwidth}{!}{\includegraphics{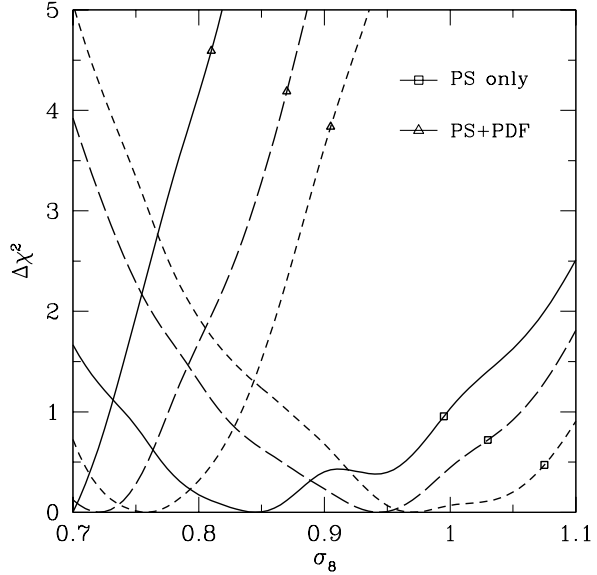}}
\caption{A comparison between the $\Delta \chi^{2}$ versus $\sigma_{8}$ 
 computed from the PS alone (curves with  squares), and
from the PS+PDF data (curves with triangles).  $\Delta\chi^2$ was
obtained after varying $\sigma_8$ and $\la F\ra$, and marginalizing
over the value of the mean flux.  The mean IGM temperature is $\thgg=1$
(solid curves), 1.5 (long-dashed) and 2 (short-dashed). The filtering
is $\kf=10\hmmpc\thgg^{-1/2}$ (cf.  Section~\S\ref{tdep}), and the
adiabatic index has a fixed value $\gamma=1.3$. The spline
interpolation was performed using the diagonal elements of the PDF
covariance matrix solely.}
\label{fig12}
\end{figure}

The RSI model matches the PS somewhat better than our scale-invariant
model with rms fluctuation $\sigma_8=0.9$, but overestimates the  PDF
in the range $F\bsim 0.6$ unless the temperature is very high,
$\thgg\bsim 2$.  A scale-invariant model with $\sigma_8=0.9$ matches
better the flux PDF but tends to overestimate $\dfk$ on scale $k\lsim
0.01\skm$ when the temperature is large, $\thgg\bsim 1.5$.  However,
the agreement with the observed flux power spectrum can be improved by
lowering the value of the normalisation amplitude, $\sigma_{8}$.  Most
importantly, models which match best the PS data alone  usually do not
yield a good fit to  the PDF.  The discrepancy worsens with decreasing
temperature. In the case of the $\sigma_8=0.9$, scale-invariant
cosmology, the disagreement with the  observed PDF is particularly
severe for $\thgg\bsim 2$. We also computed the line statistics for
some of our best fit  models. We found that they all reproduce
successfully the slope  and normalisation of the column density
distribution $f(\nhi)$, as  well as the overall line-width
distribution $f(b)$. The simulated line-width distribution is probably
affected by resolution effects. Yet the  constraints on  the
primordial power spectrum are insensitive to the details of  the
line-width statistic.  Degeneracies among the parameters diminish the
ability of the data to constrain the shape and normalisation of the
primordial  power spectrum (e.g. Zaldarriaga, Hui \& Tegmark 2001;
Zaldarriaga, Scoccimarro \& Hui 2003). In this respect, we have found
that the pairs  $\gamma$ - $\kf$ and $\thg$ - $\sigma_8$ are
degenerate when we marginalize  over the mean transmitted flux. To
proceed further, we have assumed  that the filtering wavenumber $\kf$
is related to the IGM temperature according to
$\kf\propto\thgg^{-1/2}$. The reasons which motivate this choice are
discussed in Appendix~\S\ref{filtering}.  Under that assumption, a
chi-squared minimization indicates that the normalisation amplitude is
likely to be $0.7\lsim\sigma_8\lsim 0.9$ for a reasonable reionization
scenario and temperatures $1\lsim\thgg\lsim 2$. For   $\thgg\bsim
1.5$, the data favour models with normalisation  $\sigma_8\lsim 0.8$
rather than models with $\sigma_8\bsim 0.8$, the later slightly
overestimating $\dfk$ in the range $k\lsim 0.01\skm$.  It should also
be noted that taking into account the full error matrix  of the flux
PDF systematically lowers the best fit values of $\sigma_8$  by a few
percents.

Our results  suggest that the constraints inferred from measurements
of the flux power spectrum alone might be  biased. Most previous
estimates  of  $\sigma_{8}$ from \op absorption measurement relied on
the flux PS alone (e.g. Croft \etal 1999,  2002; McDonald \etal 2000,
2004b; Viel, Haehnelt \& Springel 2004; see also Viel, Weller \&
Haehnelt 2004; Seljak \etal 2004.  In  light of our results however,
it is unclear whether their best fit  models also match the PDF of the
flux (and other statistics of the  \op forest as well). To emphasize
the importance of combining several  statistics  we compare in
Fig.~\ref{fig12} the constraints on $\sigma_{8}$  obtained from the PS
data alone (curves with square) to those from from the whole PS+PDF
data set (curves with triangles). We  assumed a  filtering wavenumber
of $\kf=10\thgg^{-1/2}\hmmpc$.  The curves $\Delta\chi^2$ were
calculated for a mean IGM temperature $\thgg=1$ (solid curves), 1.5
(long-dashed) and 2 (short-dashed), but for a fixed adiabatic index
$\gamma=1.3$. Note that the best fit values of $\sigma_8$ in the case
of the PS data only are consistent with those inferred in previous
works.  Fig.~\ref{fig12} clearly demonstrates that the inclusion of
the PDF in the minimization has a large impact on the normalisation
amplitude, decreasing the best fit value of $\sigma_8$ by about 20\%.
Consequently, the analysis of Viel, Haehnelt \& Springel (2004)  and
McDonald \etal (2004b) may  overestimate the  normalisation amplitude.
Furthermore, including the PDF significantly improves the  confidence
on $\sigma_8$ as we can see from the shape of $\Delta\chi^2$.
Finally, as we discussed in Section~\S\ref{errors}, it  also
noticeably reduces the sensitivity of the results to the mean flux
level $\la F\ra$.  A joint fit of the flux power spectrum and PDF can
thus greatly improve over the standard analysis based on the flux
power spectrum  alone.

Although including the flux PDF tends to lower the  best fit value of
$\sigma_{8}$, overall our result of $\sigma_{8}=0.85-0.95$ is
consistent with the studies of Viel, Haehnelt \& Springel  (2004),
and McDonald \etal (2004b). Both of these studies were based on flux
measurements at several redshifts (McDonald \etal 2004b use eleven
redshift bins spanning the range $2.2\leq z\leq 4.2$). They also
relied on a more detailed  treatment of the gas physics by means of
hydrodynamical simulations. Our assumption of uniform smoothing  does
not account for the dependence of the smoothing on the local IGM
density and temperature.  Yet the exact relation  between the gas and
dark matter depends on  the rather poorly constrained thermal and
reionization history of the universe.  Simulations incorporating a
wider variety of  reionization histories are needed for a better
understanding of the systematics in inferring the cosmological
parameters from the  forest.   It is also unclear whether the
currently available hydrodynamical simulations give  reasonable flux
PDFs.

We have demonstrated that a joint analysis of the forest can
potentially  yield more robust estimates of the cosmological
parameters than  those inferred from the flux power spectrum alone.
Here we have included mainly the PDF and the flux power spectrum. But
other statistical measures can also be used.  The flux bispectrum
(e.g. Mandelbaum \etal  2003) offers a promising possibility as it
should be less sensitive to systematics than the PDF.

\section{acknowledgement}

We thank Patrick McDonald and our referee, Joop Schaye, for many
valuable comments on an earlier version of this manuscript. We
acknowledge stimulating discussions with Martin Haehnelt, Michael
Rauch, Ravi Sheth,  David Tytler, and Matteo Viel. This Research was
supported by the German Israeli Foundation for Scientific Research and
Development, the EC RTN network ``The Physics of the Intergalactic
Medium'', and the United States-Israel Bi-national Science Foundation
(grant \# 2002352).  VD would like to acknowledge the Institute of
Astronomy (Cambridge) and The University of Pittsburgh where part of
this work was accomplished.

\appendix

\section{The filtering length}
\label{filtering}

Gas pressure smoothes the gas distribution relative to that of the
dark matter. This effect becomes important below a (comoving) scale
$\xj=1/\kj$, the Jeans scale, which is defined as (e.g. Bi, B\"orner \&
Chu 1992)
\begin{equation}
\xj=\frac{1}{aH}\sqrt{\frac{2\gamma
k_B\thg}{3\mu m_p}}=0.176\hmpc
\left(\frac{\gamma\thgg}{1+z}\right)^{1/2}\;,
\label{app1}
\end{equation}  
where $a$ is the scale factor, $H$ the Hubble constant, $k_B$ the
Boltzmann constant and $\mu$ the mean molecular weight. Here, $\gamma$
describes the temperature-density relation,
$\tg=\thg(1+\dgnl)^{\gamma-1}$ The numerical estimate was obtained for
a $\Lambda$CDM cosmology with  matter content $\om{m}=0.3$, assuming
$\mu=0.59$, a value appropriate for  a fully ionized plasma of
primordial abundance. We should emphasize that  eq.~(\ref{app1}),
which defines the Jeans scale $\xj(t)$ in term of  physical quantities
evaluated at mean density, merely defines a characteristic scale.  The
filtering  scale $\xf(\vx,t)$ over which the IGM is  smoothed will
generally differ from the Jeans scale. For the moment, let us
consider the filtering length {\it at mean density}, $1/\khf$, which
is also a function of time alone. The redshift  evolution of $\khf$ is
complex, and depends on the details of the reionization scenario, i.e.
on the whole time evolution of $\kj$.  However, for some particular
choices of reionization history (e.g. sudden  reionization), it is
possible to work out analytic solutions to the  {\it linear} equation
governing the evolution of baryonic and dark matter in a EdS Universe
(e.g.  Peebles 1984; Bi, B\"orner \& Chu 1992; Gnedin \& Hui 1998;
Nusser 2000;  Matarrese \& Mohayaee 2002).  Under these assumptions,
Gnedin \& Hui (1998) pointed out that, at redshift  $z=3$, the
filtering length (at mean density)  is $\khf=\eta\kj$, with $\eta\sim
1.5-2.5$ for realistic reionization   scenarios. These values should
nonetheless not be taken too seriously given the uncertainties in the
relation between baryons and dark matter. Taking $\gamma=1.3$ and
$\thgg=1$ yields  $\khf\approx 7\eta\hmmpc$. Note that we divided
$\khf$ by $\sqrt{2}$ to  account for our definition of the filter,
$W=\exp(-k^2/2\kf^2)$, which reduces to $\approx 1-k^2/2\kf^2$ in the
limit of large wavenumber.

In our analysis, we smooth the Fourier modes of the dark matter density
field with a uniform Gaussian filter to obtain the  gas density and
velocity fields.  In reality, we expect the filtering length
$\xf(\vx,t) $ to be a function  of space and  time through the local
temperature and density. To get some idea of this dependence, let us
consider a spherical perturbation of physical radius $r=\xf/(1+z)$. On
the one hand, the excess pressure will try to smooth  out the
perturbation on a timescale $r/c_S$, where $c_S$ is the local speed of
sound. One the other hand, the enhanced density gives rise to an extra
inward gravitational force (mainly due to dark matter) which  tends to
increase the matter content of the perturbation on a timescale
$1/\sqrt{G\rho^{\rm th}_{\rm m}}$, where $\rho^{\rm th}_{\rm m}$ is
the dark matter density smoothed with a top-hat window of comoving
width $\xf$. Equating these two timescales allows us to express $\xf$
as
\begin{equation}
\xf=\frac{c_S}{\sqrt{G\rho^{\rm th}_{\rm m}}}(1+z)\;.
\label{app2}
\end{equation}
The  density $\rho^{\rm th}_{\rm m}$ is not equal to the gas density. 
However, both are tightly related and have similar dependence on $\xf$. 
Hence, if we assume that $\rho^{\rm th}_{\rm m}\propto\rho_{\rm g}$, and 
that the gas temperature and density follow the power-law relation 
discussed above, we have
\begin{equation}
\xf\!(z,\dgnl)=\xhf\!(z)\left(1+\dgnl\right)^{\gamma/2-1}\;,
\label{app3}
\end{equation}
where $\xhf(z)$ is the filtering length at mean density.  For an
adiabatic  index $1\lsim\gamma\lsim 1.6$, eq.~(\ref{app3}) implies
that $\xf$ depends  weakly on the gas density contrast
$\dgnl$. Notwithstanding, a uniform smoothing will lead to an
overestimation (underestimation) of the gas  density in the low (high)
density regions of the simulation. We can however adjust the strength
of the $k$-space smoothing $W$, i.e. the value  of $\kf$, such that
the smoothed dark matter power spectrum is as close as possible to 
the true gas power spectrum. This is indeed the best we can do since
it is practically impossible to mimic a non-uniform smoothing in real 
space with a $k$-space kernel.
 
To determine the optimal $\kf$, we assume that the gas density can be 
obtained locally from the dark matter density as follows~:
\begin{equation}
1+\dgnl(\vx)=\int\!\!\frac{\dd^3y}{(2\pi)^{3/2}\xf^3}
\left[1+\dmnl(\vy)\right]e^{-\frac{(\vx-\vy)^2}{2\xf^2}}\;.
\label{app5}
\end{equation}
In principle, the filtering length $\xf=\xf(\dgnl)$ should depend on
the  local gas density, and eq.~(\ref{app5}) would implicitly define
$\dgnl$.  For simplification however, we will assume that
$\xf=\xf(\dmnl)= \xhf(1+\dmnl)^{\gamma/2-1}$
(equation~\ref{app3}). The Fourier  transform of the gas density
field is simply given by
\begin{equation}
\dgnl(\vk)+\delta_{\rm D}(\vk)=
\int\!\!\frac{\dd^3x}{(2\pi)^3}\left[1+\dmnl(\vx)\right]
e^{-\frac{1}{2}k^2\xf^2}e^{-i\vk\cdot\vx}\;,
\label{app6}
\end{equation}
where $\delta_{\rm D}(\vk)$ is the Dirac delta function. The gas power
spectrum is then (ignoring the term at $\vk=0$)
\begin{equation}
P_{\rm g}(\vk)=\int\!\!\frac{\dd^3r}{(2\pi)^3}
\la\left(1+\dmnl\right)\left(1+\dmnl'\right)
e^{-\frac{1}{2}k^2(\xf^2+\xf'^2)}\ra e^{-i\vk\cdot\vr}\;,
\label{app7}
\end{equation}
where the fields $\dmnl$, $\xf$ and $\dmnl'$, $\xf'$  are evaluated at
position $\vx$ and $\vx+\vr$ respectively. $P_{\rm g}(\vk)$ is thus
the Fourier transform of a mass-weighted filter,
\begin{equation}
{\cal Z}_{\rm g}[k,\vr]=\la\left(1+\dmnl\right)\left(1+\dmnl'\right)
e^{-\frac{1}{2}k^2(\xf^2+\xf'^2)}\ra\;.
\label{app8}
\end{equation}
To estimate $\kf$, we consider the limit of small wavenumbers
$k\rarrow 0$.  Since the integrand is weighted by $r^2$, we expect the
main contribution  to the integral to arise in the limit
$|\vr|\rarrow\infty$.  Expanding the previous equation to first order 
in $k$ and taking the limit $|\vr|\rarrow\infty$, we have
\begin{eqnarray}
{\cal Z}_{\rm g}[k,\vr]&\approx&\left(1+\xi_{\rm
m}(r)\right)-\frac{k^2}{2}
\la\left(1+\dmnl\right)\left(1+\dmnl'\right)(\xf^2+\xf'^2)\ra\nonumber
\\ &=&\left(1+\xi_{\rm m}(r)\right)\left[1-k^2 \frac{\la
\left(1+\dmnl\right)\left(1+\dmnl'\right)\xf^2\ra} {1+\xi_{\rm
m}(r)}\right] \nonumber \\ &\approx&\left(1+\xi_{\rm m}(r)\right)
\left[1-k^2\la\left(1+\dmnl\right)\xf^2\ra\right]\;,
\label{app8}
\end{eqnarray}
since the cross-terms $\la\dmnl'\xf^2\ra$ and $\la\dmnl\xf^2\dmnl'\ra$, 
and the two-point correlation $\xi_{\rm m}$ tend to zero for large 
separation. Hence, in the limit $k \rarrow 0$, the gas power spectrum 
can be expressed as
\begin{equation}
P_{\rm g}(\vk)\simeq \left(1-\frac{k^2}{\kf^2}\right)P_{\rm m}(\vk)\;,
\label{app9}
\end{equation}
where $\kf$ is an effective filtering length which differs from the
filtering length $\khf=1/\xhf$ at mean density,
\begin{equation}
\frac{1}{\kf^2}=\la\left(1+\dmnl\right)\xf^2\ra= \frac{1}{\khf^2}
\int\!\!\dd\dmnl {\cal P}(\dmnl)\left(1+\dmnl\right)^{\gamma-2}\;.
\label{app10}
\end{equation}
Here, ${\cal P}(\dmnl)$ is the one-point probability distribution of
the dark matter density field. Eq.~(\ref{app10}) shows that, in the
linear regime, $\kf$ is obtained from a mass-weighted average of the
local filtering length $\xf$.  To evaluate $\kf$, we will assume that
the dark matter PDF is well  approximated by a lognormal distribution,
\begin{equation}
{\cal P}(\dmnl)\dd\dmnl=\frac{1}{\sqrt{2\pi}\sgl}e^{-\nu^2/2\sgl^2}
\dd\nu\;.
\label{app11}
\end{equation}
with $\nu=\ln(1+\dmnl)+\sgl^2/2$, and where $\sgl$ is the linear rms
of dark matter fluctuations (Coles \& Jones 1991). Although  this
approximation is inaccurate when $\sgl\bsim 1$, it should be
adequate enough to assess whether $\kf$ is larger or smaller than 
$\khf$. The calculation yields
\begin{equation}
\kf^2=\khf^2\,\left(1+\sgnl^2\right)^{(\gamma-2)(3-\gamma)/2}\;,
\label{app12}
\end{equation}
where we have used $1+\sgnl^2=\exp(\sgl^2)$. Note that, since $\xf$
depends on the local gas density through relation~(\ref{app3}), we
should  expect $\sgnl$ to be a smoothed version of the dark matter rms
fluctuation  amplitude, i.e. a function of $\kf$. Nonetheless,
equation~(\ref{app12}) clearly shows that, for an adiabatic index
$1\lsim\gamma\lsim 1.6$,  the mass-weighted filtering wavenumber $\kf$
is smaller than that at mean density. This follows from the fact that
most  of the mass resides in moderately high-density regions where the
temperature  $\tg$ is significantly larger than $\thg$ (e.g. Bi \&
Davidsen 1997).   Consequently, given that i) $\khf\propto\kj$ and ii)
$\kf\leq\khf$,  we can place an upper limit on the filtering
$\kf$. For the particular choice $\gamma=1.3$, one has $\kf\lsim
7\eta\thgg^{-1/2}\hmmpc$. Bearing in mind that the IGM temperature is 
most probably larger than $10^4\kel$ at $z=3$ (e.g. Schaye \etal 2000), 
we believe that $\kf\lsim 14\hmmpc$ should be a reliable lower limit 
on the amount of filtering at $z=3$ for a reasonable history of the 
Universe.

\end{document}